\documentclass[a4paper]{article}
\usepackage{graphicx}
\usepackage{amsmath}
\usepackage{amssymb}
\usepackage{amsthm}
\usepackage{subfigure}
\usepackage{multicol, blindtext}

\newtheorem{thm}{Theorem}[section]
\newtheorem{cor}[thm]{Corollary}

\newtheorem{prop}[thm]{Proposition}

\newtheorem{rem}[thm]{Remark}
\numberwithin{equation}{section}
\newcommand{\norm}[1]{\left\Vert#1\right\Vert} 
\newcommand{\abs}[1]{\left\vert#1\right\vert}

\newcommand{\Real}{\mathbb R}
\newcommand{\eps}{\varepsilon}

\title{ \textbf{Optimal investment-consumption problem: post-retirement with minimum guarantee }}%

\begin{document}
\date{}%

\maketitle
\begin{center}
\author{{Hassan Dadashi}}\\
\emph{Department of Mathematics, \\
 Institute for Advanced Studies in Basic Sciences (IASBS),\\
  Zanjan, 45137-66731, Iran\\
  dadashi@iasbs.ac.ir}
\end{center}

\begin{abstract}
We study the optimal investment-consumption problem for a member of defined contribution plan during the decumulation phase. For a fixed annuitization time, to achieve higher final annuity, we consider a variable consumption rate. Moreover, to have a minimum guarantee for the final annuity, a safety level for the wealth process is considered. To solve the stochastic optimal control problem via dynamic programming, we obtain a Hamilton-Jacobi-Bellman (HJB) equation on a bounded domain. The existence and uniqueness of classical solutions are proved through the dual transformation. We apply the finite difference method to find numerical approximations of the solution of the HJB equation. 
Finally, the simulation results for the optimal investment-consumption strategies, optimal wealth process and the final annuity for different admissible ranges of consumption are given. Furthermore, by taking into account the market present value of the cash flows before and after the annuitization, we compare the outcomes of different scenarios.   
 \end{abstract}




\section{\textbf{Introduction}}
In this work, focusing on the decumulation phase in a defined contribution plan, we fix the annuitization time and investigate the optimal investment-consumption strategies in a Brownian market model with a time dependent mortality rate. In formulating the loss function, we consider a target for the consumption during the decumulation phase and a target for the terminal accumulated wealth.   
Since avoiding the ruin possibility, and sometimes having a minimum guarantee on the return, is an essential issue in pension fund management, motivated from \cite{kn:G}, we assume a minimum guarantee for the terminal wealth. 

In this paper, different from the existing literature, we assume a variable consumption rate. 
Assuming a fixed rate of consumption during the whole decumulation phase, which is usually a long period, is far from optimality. Our simulation results justify  that assuming a variable consumption rate,  although quite restricted, yields much higher final annuities. However, to cover the essential expenses of the retiree, it is reasonable to fix a lower bound for the consumption rate. Therefore, having in mind a desired rate of consumption, we assume that the consumption rate varies between the two limits $C_1$ and $C_2$, in which $C_1 > C_2$.  
To compare the optimal portfolios corresponding to different scenarios of the admissible range of consumption $[C_2, C_1]$, we take into account the market present value of the cash flows before and after the annuitization. 
                                               
 Gerrard et al. \cite{kn:Ger1} study the portfolio optimization problem post-retirement when the loss function is defined by a target function on the wealth process and the annuitization time and the consumption rate are fixed.  In \cite{kn:Ger2}, the authors violate the fixed consumption rate assumption.  In a similar framework, Di Giacinto et al. \cite{kn:G} explore the optimal investment strategy when a minimum guarantee for the final wealth is assumed and the consumption rate is fixed. They obtain the closed form for the optimal strategies just in the case that the running cost is neglected. For the general form of the loss function, a numerical algorithm for finding approximations of optimal investment strategies is developed in \cite{kn:D}. 
 Using a loss function similar to the one that is used in this work but without any constraint on the wealth process, Gerrard et al. \cite{kn:Ger3} investigate the optimal annuitization time together with the optimal consumption-investment strategies.  Vigna \cite{kn:V} shows that the target based approach in portfolio optimization of pension funds yields portfolios that are efficient in the mean-variance setting. 
 
 As an extension of our framework, the annuitization time can be considered as a control variable which can reduce to some extent the annuity risk, the risk that exists when the interest rate is high.  Furthermore, since in our problem the time horizon is long, considering a non-constant interest rate model yields a more realistic framework. 

Assuming the same constraint on the final wealth as in Di Giacinto et al. \cite{kn:G}, we  consider a variable consumption rate, as a novelty in our framework. Moreover, we consider  a running cost term in the loss function based on the consumptions. To solve the given stochastic optimal control by applying dynamic programming, we get a nonlinear HJB equation which does not seem to have an explicit solution. Using  dual transformation, we prove that the value function is sufficiently smooth to imply the verification theorem.  Then, to obtain numerical  approximations of  the value function and the optimal investment-consumption strategies, we employ a numerical scheme based on a fully implicit backward in time finite difference method. Moreover, to tackle the algebraic nonlinear systems that appear in the numerical procedure, the policy iteration method is applied. 
The policy iteration method is a well studied method in finding approximations of  solutions of optimal control problems (see \cite{kn:FL}, \cite{kn:FW} and \cite{kn:SR}) in which the value function and the strategies are derived iteratively to converge to the solution of the corresponding HJB equation and the optimal strategies. In \cite{kn:CD}, this method is employed in a portfolio selection problem with transaction costs. 
   
The portfolio optimization problem post-retirement has been studied by many scholars who consider different utility or loss functions, different control variables and different wealth dynamics. Furthermore, different constraints  on the control variables or on the wealth process can be assumed. We mention here a few related works. Milevsky and Young \cite{kn:MY} study extensively the optimal annuitization and investment-consumption problem for a time dependent mortality rate in the\emph{ all or nothing} market and also the more general \emph{anything anytime}s market, where gradual annuitization strategies are allowed. Milevsky et al. \cite{kn:MMY} derive the optimal investment and annuitization strategies for a retiree whose objective is to minimize the ruin probability when the consumption rate is fixed. Stabile \cite{kn:S} studies the optimal investment-consumption problem and investigates the optimal time for purchasing the annuity subject to a constant force of mortality by considering different utility functions based on the consumption and the final annuity. Blake et al. \cite{kn:BCD} compare the immediate annuitization at the retirement time with distribution programs involving differing exposures to equities during the retirement. Albrecht and Maurer \cite{kn:AM} compare the immediate annuitization and the income drawdown option and determine the probability of running out of money before an uncertain date of death. 

The article is organized as follows. In the next section, we specify the market model and the restrictions of control variables. In the third section, the set of admissible strategies, the loss function and also the value function are determined. In addition, some properties and the domain of the wealth process are identified. In Section 4, the corresponding HJB equation is written and, using  dual transformation, the regularity of the solution is proved. In Section 5, we express the numerical algorithm to find approximations of the solution of the HJB equation. Finally in Section 6, the simulation results for the final annuity, optimal investment-consumption strategies and optimal wealth process are presented. Furthermore, to compare the outcomes of different scenarios, the present values of the corresponding cash flows are revealed. 

\section{The Market Model}
We consider a Brownian market model that consists of a risky and a risk-less asset with the dynamics: 
\begin{align}
&dP_t=P_t(\mu dt+\sigma dB_t), \nonumber\\
&dA_t=rA_tdt,\nonumber
\end{align}
where $B(\cdot)$ is a Brownian motion on the filtered probability space $(\Omega, \mathcal{F}, \mathbb{F}, \mathbf{P})$ and $r$ is the fixed interest rate. So, the risky asset is a geometric Brownian motion with constant volatility $\sigma$ and expected return $\mu=r+\sigma \beta$, in which $\beta $ is its   Sharpe ratio.

At any time $t$, let $\pi^1_t$ and $1-\pi^1_t$ be the fractions of the fund's portfolio that are invested in the risky and in the risk-less asset, respectively.  Denoting the consumption rate at this time by $\pi^2_t$, we have the following dynamics of the wealth process (or the fund value dynamics) 
\begin{align}
\begin{cases}
dX_t=\{[\pi^1_t(\mu-r)+r]X_t-\pi^2_t \} dt+\sigma \pi^1_t X_t dB_t,\label{wealtheq}\\
X_0=x_0. 
\end{cases}
\end{align}

We restrict the consumption rate to the interval $[C_2, C_1]$, in which $C_1$ is the desired rate of consumption and usually is set to be equal to the annuity that is purchasable by the accumulated wealth at the retirement time and $C_2$ is set as  the minimum admissible  consumption rate during the decumulation phase. Furthermore, the investment control variable $\pi^1$ is restricted to the interval $[0, L]$. This means that the short-selling is prohibited. But, we can borrow from the money market up to $L$ times the fund value. 

Our market model can be extended in several directions. Among many other models, Boulier  et al. \cite{kn:BHT} consider a market model with stochastic interest rate. Han and Hung \cite{kn:HH} equip the model with the inflation rate which has important consequences on the optimal strategies. Deelstra et al. \cite{kn:DGK}, in a stochastic interest rate framework, consider a market model that consists of three assets, a risky asset, a risk-less asset and a bond. Hainaut and Deelstra \cite{kn:HD} investigate the optimal time for the annuity purchase when the risky asset has the jump-diffusion dynamics and the economic utility function is replaced by the  expected present value operator.  

\section{Optimal Control Problem} 
Let the decumulation phase be represented by the time interval $[0,T]$. We consider a retiree with age $a_0$ who is going to postpone the annuitization until the age $a_1=a_0+T$. The main aim of a retiree in postponing the annuity purchase is to reach the desired annuity. Let $F$ be the target for the terminal accumulated wealth, $X_T$, by which the retiree can purchase the desired annuity at the age $a_1$. In other words, if $a_{a_1}$ is the actuarial value of the unitary lifetime annuity at the age $a_1$, then $\frac{ F}{a_{a_1}}$ would be the desired annuity. Furthermore, during the decumulation phase, part of the  retiree's concern is  on the consumption and he or she would like to consume at the maximum admissible rate  $C_1$.  
Therefore, we write the loss function by two terms. One term, as the running cost, is defined as the penalizing deviation of the interim  consumption from the desired rate $C_1$ and is weighted  by $\kappa > 0$. The second term is written as the penalizing deviation from the targeted annuity $\frac{ F}{a_{a_1}}$ at the terminal time. 

We assume that the mortality rate $\nu_u$, $u\geq a_0$, is independent of the asset dynamics. So, denoting $\eta_t=e^{-\int^t_0 (r+\nu_{s+a_0})ds}$,  the loss function is written as 
\begin{align}
\kappa \int^T_0  \eta_t (C_1-\pi^2_t)^2dt+ \eta_T \left(\frac{ F-X_T}{a_{a_1}}\right)^2, \label{lossfunc}
\end{align}
where $X_T$ is the value of the solution of Equation (\ref{wealtheq}) at the terminal time. 

For any $0\leq t\leq T$, consider the filtration $\mathbb{F}^t:=(\mathcal{F}^t_s)_{s\in [t,T]}$ where $\mathcal{F}^t_s$ is the $\sigma$-algebra generated by the random variables $(B_u-B_t)_{u\in [t,s]}$. 
Then the strategies $\pi^1(\cdot)$ and $\pi^2(\cdot)$, that are supposed to be $\mathbb{F}^t$-progressively measurable processes, are chosen from  $\mathcal{L}^2(\Omega \times [t,T]; [0, L])$ and $ \mathcal{L}^2(\Omega \times [t,T]; [C_2, C_1])$, respectively. It should be noted that for the controls that are chosen from these classes, the following equation has a unique strong solution; see \cite[Section 5.6.C]{kn:KS}, 
\begin{align}
\begin{cases}
dX_s=\{[\pi^1_s(\mu-r)+r]X_s-\pi^2_s \} ds+\sigma \pi^1_s X_t dB_s,  \qquad s\geq t,\label{wealtheq2}\\
X_t=x. 
\end{cases}
\end{align}
which is denoted by $X(\cdot ; t, x, \pi^1(\cdot), \pi^2(\cdot)).$ 

To apply dynamic programming in finding the optimal strategies which minimize the above loss function, we have to embed the stochastic  optimal control problem in a family of problems with varying initial time.  
So, for any initial time $0\leq t\leq T$ and initial wealth $x>0$, a stochastic optimal control problem is considered, by minimizing the following loss function over the set of admissible strategies that are specified in (\ref{adm}), 
\begin{align}
\kappa \int^T_t  \eta_s (C_1-\pi^2_s)^2ds+ \eta_T \left(\frac{ F-X(T; t, x, \pi^1(\cdot), \pi^2(\cdot))}{a_{a_1}}\right)^2. \label{lossfunc1}
\end{align}

The main feature of this work's framework is assuming  a minimum guarantee for the final annuity or equivalently for the terminal wealth. Let the safety level for the terminal wealth amount be $S$, which  depends on the retiree's level of risk aversion. 
So, the set of admissible  strategies at time $0\leq t\leq T$, when $X_t=x$, reduces to 
\begin{align}
	\tilde{\Pi}_{ad}(t,x):=\{&\pi^1\in \mathcal{L}^2(\Omega \times [t,T]; [0, L]), \pi^2\in \mathcal{L}^2(\Omega \times [t,T]; [C_1, C_2]) | \; \pi^1, \pi^2 are\nonumber\\
	 & \mathbb{F}^t- prog.\; meas., X(T; t,x,\pi^1(\cdot), \pi^2(\cdot)) \geq S \; a.s. \}.\label{adm} 
\end{align}

Regarding the loss function (\ref{lossfunc1}), for any $(t,x)\in [0,T]\times \Real^+$,  the following objective functional on the set of admissible strategies $ \tilde{\Pi}_{ad}(t,x)$ is defined 
\begin{align}
\tilde{J}(t,x;\pi^1(\cdot), \pi^2(\cdot)):=\mathbb{E}^x &[\kappa \int^T_t \eta_s (C_1-\pi^2_s)^2ds\nonumber\\
&+ \eta_T \left(\frac{F-X(T; t,x,\pi^1(\cdot), \pi^2(\cdot))}{a_{a_1}}\right)^2 ],\label{oldfunc} 
\end{align}
where  $\mathbb{E}^x$ stands for the expectation subject to $X_t=x$. 
Our goal is to find the admissible strategies that minimize the above functional. To solve this stochastic optimal control problem via  dynamic programming, the following value function is defined  
\begin{align}
\tilde{V}(t,x):=\inf_{\pi^1(\cdot), \pi^2(\cdot)\in \tilde{\Pi}_{ad}(t,x)} \tilde{J}(t,x;\pi^1(\cdot), \pi^2(\cdot)).\label{value.f}
\end{align}
Moreover, the corresponding HJB equation is specified in the next section.

Now, using the loss function definition, dynamic programming and the safety level constraint, we explore some properties and the domain of the state process. 
The minimum guarantee constraint at the terminal time imposes consequently a constraint on the wealth process during the decumulation phase. Actually, the following curve is as a  barrier for the wealth process, 
\begin{align}
S(t)=\frac{C_2}{r}-(\frac{C_2}{r}-S)e^{-r(T-t)}, \qquad 0\leq t\leq T. \label{safe}
\end{align}
Moreover, the loss function formulation, (\ref{lossfunc1}), indicates that, at any time $t\in [0,T]$, the wealth amount 
\begin{align}
F(t)=\frac{C_1}{r}+(F-\frac{C_1}{r})e^{-r(T-t)}, \label{targfun}
\end{align}
guarantees reaching the target $F$ at the terminal time T, by investing the whole portfolio during the time interval $[t,T]$ in the risk-less asset and consuming at the maximum rate $C_1$.  Here, we call the function  $F(\cdot)$ the target function. 

The above statements are proved precisely in the following proposition. 

\begin{thm}\label{absorb}

Let the initial wealth $x$ of the stochastic optimal control problem with initial time $t$ and loss function (\ref{lossfunc1}) lies in the interval $[S(t), F(t)]$. If at any time $s \in [t, T ]$, the wealth reaches the boundary of the interval $[S(s),F(s)]$, then, from that $s$ on, the optimal wealth amount remains in such boundary till time $T$. We call this property of the boundaries the absorbing property. \\

\begin{proof}
 The target function formulation (\ref{targfun}), implies 
$$dF(s)=(rF(s)-C_1)ds.$$
Moreover, notice that  taking the null investment strategy, $\pi^1\equiv 0$, and the maximum consumption strategy, $\pi^2\equiv C_1$, yield the same dynamics for the wealth process 
$$dX_s=(rX_s-C_1)ds. $$
Suppose that, by applying the optimal strategy $\pi=(\pi^1, \pi^2)$ at the point $(t,x)$, the wealth process  hits the target $F(s)$ at some time $s\in [t,T]$.  The above observation indicates that the strategy $(0, C_1)$ at the point $(s, F(s))$ yields zero cost, and therefore it is optimal. On the other hand, due to the dynamic programming principle, the restriction of the optimal strategy  $\pi=(\pi^1, \pi^2)$ to $[s,T]$ gives the optimal strategy at the point $(s,F(s))$, which is $(0,C_1)$. So, after the time $s$, the optimal wealth process $X(\cdot; t,x, \pi^1(\cdot), \pi^2(\cdot))$ will remain on the curve $\{F(u), s\leq u\leq T \}$ and never overtakes the upper barrier.

Now, assume that the wealth amount be equal to $S(s)$ at a time $s\in [t, T]$, $X_s=S(s)$. Then by applying the strategy $\pi=(0,C_2)$, we have $dX_u=dS(u)$, for $u\geq s$. So, this strategy keeps the wealth process on the curve $\{S(u), s\leq u\leq T\}$. We prove that this is the only admissible strategy at the point $(s, S(s))$. 

For any admissible investment strategy $\pi^1(\cdot)$, the process 
$$Y_u=X(u; s,S(s),\pi^1(\cdot), C_2), \quad u\geq s,$$ 
has the following dynamics under the risk-neutral measure, the measure that makes the expected return of the risky asset equal to the interest rate $r$, 
$$dY_u=(rY_u-C_2)du+\pi^1_u Y_udW_u, $$
in which $W_u=B_u+\frac{\mu-r}{\sigma} u$ stands for the Brownian motion under this measure.  

Therefor, letting $Z_u=Y_u-S(u)$, we have $Z_s=0$ and  for $s< u< T$
$$dZ_u=rZ_u du+\pi^1_u(Z_u+S_u)dW_u,$$ 
which indicates that the discounted value of $Z_u$, $\tilde{Z}_u=e^{-ru}Z_u$, is a martingale under the risk-neutral measure. So, we have $\tilde{\mathbb{E}}(\tilde{Z}_T)=\tilde{\mathbb{E}}(\tilde{Z}_s)=0$, in which $\tilde{\mathbb{E}}$ denotes the expectation operator under this measure. This equality indicates 
\begin{align}
\tilde{\mathbb{E}} X(T; s,S(s),\pi^1(\cdot), C_2) =S(T)=S. \label{p1}
\end{align}
On the other hand, the minimum guarantee constraint imposes the following inequality under the physical and equivalently under the risk-neutral measure   
 \begin{align}
 X(T; s,S(s),\pi^1(\cdot), C_2) \geq S, \quad a.s. \label{p2}
 \end{align}
Furthermore, any investment strategy other than the null  strategy yields a non-constant random terminal wealth. So, due to (\ref{p1}) and (\ref{p2}), when $\pi^2\equiv C_2$,  the only admissible investment strategy is  $\pi^1\equiv 0$. 

Moreover, the higher consumption rate the less terminal wealth amount. This means that the only admissible strategy, and therefore optimal  strategy, is $\pi=(0, C_2)$. 
\end{proof}
\end{thm}

\begin{rem}
If $X_0 > F(0)$, it is clear from the first part of the above argument that the strategy $\pi=(0, C_1)$ yields the desired consumption and a terminal wealth that is greater than the target $F$. However, due to the second term of the loss function (\ref{lossfunc}), the loss function increases when the terminal wealth increases in the region $[F, +\infty)$. Therefore, in the case $X_0 > F(0)$, this loss function, and consequently the corresponding optimal control problem, is not meaningful.   

If $X_0 < S(0)$, we conclude from the second part of the argument that the constraint on the terminal wealth does not hold for any admissible strategy. So, the problem does not have a solution. 
\end{rem}

\begin{cor}
The absorbing property of the lower border indicates the following equivalent representation for the admissible strategies 
\begin{align}
	\tilde{\Pi}_{ad}(t,x):=\{&\pi^1\in \mathcal{L}^2(\Omega \times [t,T]; [0, L]), \pi^2\in \mathcal{L}^2(\Omega \times [t,T]; [C_1, C_2])| \pi^1, \pi^2\; are \nonumber\\
	 & \mathbb{F}^t- prog.\; meas., X(s; t,x,\pi^1(\cdot), \pi^2(\cdot)) \geq S(s),\; t\leq s\leq T \; a.s. \}.\label{adm2}
\end{align}

\end{cor}

\section{\textbf{The HJB Equation}}

Due to the definition of the value function $\tilde{V}$, (\ref{value.f}), the Bellman principle yields the following HJB equation on the domain $\mathcal{C}:=\{(t,x) | t\in[0,T], S(t)\leq x\leq F(t) \}$; see \cite[Chapter 11]{kn:O}, 

\begin{align}
\inf_{(\pi^1, \pi^2)\in [0, L]\times [C_2, C_1]}\{ \frac{\partial \tilde{V}(t,x)}{\partial t}&+ \tilde{\mathcal{A}}\tilde{V}(t,x)+ \kappa \eta_t (C_1-\pi^2)^2 \}=0, \label{main1}
\end{align}
where $\tilde{\mathcal{A}}$ is the generator of the diffusion process $X$, given in (\ref{wealtheq2}),  
\begin{align}
\tilde{\mathcal{A}} =\{(\pi^1[\mu-r]+r)x-\pi^2\} \frac{\partial }{\partial x}+ \frac{1}{2} \sigma^2 (\pi^1)^2 x^2 \frac{\partial^2 }{\partial x^2}.\nonumber
\end{align}
Additionally, the definition of $\tilde{V}$ and the absorbing property of the upper and lower borders imply the following boundary conditions
\begin{align}
&(i) \tilde{V}(T,x)= \eta_T \left(\frac{F-x}{a_{a_1}}\right)^2, \qquad \qquad \qquad \qquad \qquad \qquad \quad   x\in [S,F],\nonumber\\
&(ii)\tilde{V}(t,S(t))=\eta_T \left(\frac{ F-S}{a_{a_1}}\right)^2 +\kappa \int^T_t \eta_s ds \left(C_1-C_2\right )^2, \quad  t\in [0,T],\label{mainboundary}\\
&(iii)\tilde{V}(t,F(t))=0,\qquad \qquad  \qquad \qquad \qquad \qquad \qquad \qquad \quad  t\in [0,T].\nonumber
\end{align}

\subsection{Reducing the domain to a rectangle}
The domain of Equation (\ref{main1}) is irregular, the upper and the lower borders are curved.  Since we are going to apply the finite difference method as part of our algorithm, we apply a change of variable that converts the domain to a rectangle. To this end, we define the diffeomorphism  $\mathcal{L}: \mathcal{C} \rightarrow \mathcal{C}^{\prime}$, where $\mathcal{C}^{\prime}:=\{(t,z) | t\in[0,T], S\leq z\leq F \}$  and 

\begin{align}
(t,x)\rightarrow (t,z)&=\mathcal{L}(t,x)=(t,\mathcal{L}_1(t,x))\nonumber\\
&:=\left(t,xe^{r(T-t)}+\left[C_1+(C_2-C_1)\frac{F(t)-x}{F(t)-S(t)} \right]\frac{1-e^{r(T-t)}}{r}\right).\label{diffeo}
\end{align}
Notice that for all $0\leq t\leq T$
\begin{align}
\mathcal{L}_1(t,S(t))=S,\qquad \qquad \mathcal{L}_1(t,F(t))=F. \label{borders}
\end{align}
Then, we define the diffusion process $Z$ as 
$$Z_t:=\mathcal{L}_1(t,X_t)=X_tG(t)+H(t),\qquad  \qquad t\in [0,T], $$
in which 
\begin{align}
G(t)=e^{r(T-t)}+ \frac{C_1-C_2}{F(t)-S(t)} \frac{1-e^{r(T-t)}}{r}=\frac{F-S}{F(t)-S(t)},\label{Gfunc}
\end{align}
\begin{align}
H(t)=\left(C_1+ \frac{F(t)(C_2-C_1)}{F(t)-S(t)}\right) \frac{1-e^{r(T-t)}}{r}= \frac{C_2F-C_1S}{F(t)-S(t)} \frac{e^{-r(T-t)}-1}{r}.\nonumber
\end{align}

\begin{prop}
The process $Z$ satisfies the following dynamics
\begin{align}
dZ_t=&X_t\left[-\frac{(F-S)(C_2-C_1+r(F-S))e^{-r(T-t)} }{(F(t)-S(t))^2} \right]dt\nonumber\\
&+G(t) \left\{[(\pi^1_t(\mu-r)+r)X_t-\pi^2_t ]dt+\sigma \pi^1_tX_tdW_t \right \}\nonumber\\
&+\left[\frac{C_2F-C_1S}{F(t)-S(t)}\right]e^{-r(T-t)}dt\nonumber\\
&+\frac{(C_2F-C_1S)(C_2-C_1+r(F-S))}{(F(t)-S(t))^2}\frac{e^{-r(T-t)}-e^{-2r(T-t)}}{r} dt.\label{zproc}
\end{align}
\begin{proof}
Due to the product rule 
$dZ_t=X_tdG(t)+G(t)dX_t+dH(t) $
and regarding the dynamics of $X$, (\ref{wealtheq}), we get directly  the above dynamics.  
\end{proof}

\end{prop}

Notice that by a few manipulations we get
$$X_t=\frac{r[F(t)-S(t)]Z_t-[C_2F-C_1S](e^{-r(T-t)}-1) }{r(F-S)}.$$
Therefore, defining the function
\begin{align}
K(t,z):=&\frac{[F(t)-S(t)]}{F-S}z-\frac{[C_2F-C_1S](e^{-r(T-t)}-1) }{r(F-S)} \nonumber\\
=&K_1(t)z-K_2(t),\label{K}  
\end{align}
we can rewrite the dynamics (\ref{zproc}) as 
\begin{align}
dZ_t=&\left\{K(t,Z_t) G^{\prime}(t)+G(t) [(\pi^1_t(\mu-r)+r)K(t,Z_t)-\pi^2_t ]+H^{\prime}(t)\right\}dt \nonumber\\
&+G(t)\sigma\pi^1_t K(t,Z_t) dW_t.\label{zproc2}
\end{align}

\begin{prop}
For each $(t,z)\in \mathcal{C}^{\prime}$, the set of admissible strategies analogous to (\ref{adm2}), can be written as  
\begin{align}
\Pi_{ad}(t,z)=\{&\pi^1\in \mathcal{L}^2(\Omega \times [t,T]; [0, L]), \pi^2\in \mathcal{L}^2(\Omega \times [t,T]; [C_2, C_1])| \pi^1, \pi^2\; are \nonumber\\  
&  \mathbb{F}^t- prog. meas. , Z(s;t,z,\pi^1(\cdot), \pi^2(\cdot))\geq S,\;  t\leq s\leq T \; a.s. \},\nonumber
\end{align}
where we define 
$$Z(s;t,z,\pi^1(\cdot), \pi^2(\cdot)):=\mathcal{L}_1(s,X(s;t,x,\pi^1(\cdot), \pi^2(\cdot))), \quad t\leq s\leq T,$$ 
in which $x=\mathcal{L}^{-1}_1(t,\cdot)(z)$.

\begin{proof}
This representation is obtained due to relations in (\ref{borders}) and the absorbing property of upper and lower borders of the domain $\mathcal{C}$.
\end{proof}

\end{prop}

 We should emphasize that there is the equality $\Pi_{ad}(t,z)=\tilde{\Pi}_{ad}(t,x)$. 

Because of the above change of variable, we should reformulate our stochastic optimal control problem on the new domain using the process $Z$.  Analogous to (\ref{oldfunc}), we define for any $(t,z)\in \mathcal{C}^{\prime}$ the following functional  on $\Pi_{ad}(t,z)$,  
\begin{align} 
J(t,z,\pi^1(\cdot), \pi^2(\cdot)):=\mathbb{E}^z [&\kappa\int^T_t \eta_s (C_1-\pi^2_s)^2ds\nonumber\\
&+\eta_T \left(\frac{F-Z(T;t,z,\pi^1(\cdot), \pi^2(\cdot))}{a_{a_1}}\right)^2 ]. \label{newJ}
\end{align}
Then, defining the value function $V$ as 
\begin{align}
V(t,z):=\inf_{\pi^1(\cdot), \pi^2(\cdot)\in\Pi_{ad}(t,z)} J(t,z;\pi^1(\cdot), \pi^2(\cdot)), \qquad (t,z)\in \mathcal{C}^{\prime},\label{value1}
\end{align}
we get the following HJB equation on the domain $\mathcal{C}^{\prime}$ 
\begin{align}
\inf_{(\pi^1, \pi^2)\in [0, L]\times [C_2, C_1]}\left\{ \frac{\partial V(t,z)}{\partial t}+ \mathcal{A}V(t,z) +\kappa \eta_t (C_1-\pi^2)^2\right \}=0, \label{maineq}
\end{align}
where $\mathcal{A}$, the generator of the diffusion process $Z$, is written, due to the dynamics (\ref{zproc2}), as 
\begin{align}
\mathcal{A}=&\alpha(t,z)\frac{\partial }{\partial z}+\beta(t,z)\frac{\partial^2 }{\partial z^2}:= \nonumber\\
&\left\{K(t,z) G^{\prime}(t)+G(t) [(\pi^1(\mu-r)+r)K(t,z)-\pi^2 ]+H^{\prime}(t)\right\}  \frac{\partial }{\partial z}\nonumber\\
&+\frac{1}{2}G^2(t)\sigma^2(\pi^1)^2 K^2(t,z) \frac{\partial^2 }{\partial z^2}.\label{diffA}
\end{align}

Since before the annuitization the loss function just depends on the consumptions, the  boundary conditions (\ref{mainboundary}) are concluded similarly for $V$. Moreover, in characterizing the dual transformation, the Neumann boundary condition at $z=F$, that is proved in Prop. \ref{neuman}, is needed. So, the boundary conditions of $V$ are written as  
\begin{align}
&(i)\; V(T,z)= \eta_T \left(\frac{F-z}{a_{a_1}}\right)^2,\qquad  \qquad \qquad \qquad  \qquad \qquad\quad\;\;\; z\in [S,F],\nonumber\\
&(ii)\; V(t,S)=\eta_T \left(\frac{F-S}{a_{a_1}}\right)^2+\kappa (C_1-C_2)^2 \int^T_t \eta_sds, \quad\qquad\;  t\in [0,T],\nonumber\\
&(iii) \; V_z(t,F)=0,  \qquad \qquad\qquad \qquad \qquad \qquad \qquad \quad\qquad \qquad t\in [0,T],       \label{boundary}\\
&(iii^{\prime})V(t,F)=0, \qquad \qquad\qquad \qquad \qquad \qquad \qquad \quad\qquad \qquad t\in [0,T]. \nonumber
\end{align}
\begin{prop}\label{neuman}
For any $t\in[0,T]$, the value function has left derivative at $z=F$ and it is equal to zero, $V_z(t,F)=0$.
\begin{proof}
Let $z\in (S,F)$ be the initial state. To find the result of applying the strategy $\pi= (\pi^1, \pi^2)= (0, C_3)$, in which $C_3=C_1+(C_2-C_1)\frac{F-z}{F-S}$, to the wealth process (\ref{zproc2}), we apply this  strategy to (\ref{wealtheq2}) and then use the transformation $\mathcal{L}_1$. Let $x=[\mathcal{L}_1(t,\cdot)]^{-1}(z)$ be the corresponding initial state in the domain $\mathcal{C}$. Denoting $X_s=X(s;t,x,\pi^1(\cdot), \pi^2(\cdot))$, we have $X_t=x$ and 
$$X_s=\frac{C_3}{r}+(z-\frac{C_3}{r})e^{-r(T-s)}, \quad s\geq t. $$ 
Like to the equality (\ref{C_3}), we can conclude that the process $Z_s=\mathcal{L}_1(s,, X_s)$ does not move from the level $z$, and in particular $Z_T=z$. 

Now, the inequality $V(t,z)\leq J(t,z,0, C_3)$ and the relation $V(t,F)=0$ imply that  
\begin{align}
-\frac{\kappa(C_1-C_3)^2\int^T_t\eta_sds+\eta_T(\frac{F-z}{a_{a_1}})^2 }{F-z}\leq \frac{V(t,F)-V(t,z)}{F-z}\leq 0.\nonumber
\end{align}
Regarding the definition of $C_3$, the left hand side is rewritten as 
$$(F-z)\left[\kappa(\frac{C_2-C_1}{F-S})^2\int^T_t\eta_sds+\frac{\eta_T}{a^2_{a_1}}\right],$$
which tends to zero, as $z\uparrow F$, and concludes the claim. 
\end{proof}
\end{prop}

 In the following proposition, we state that the strategies that are optimal in the new optimal control problem, that has been formulated after the change of variable, are  optimal in the original problem too. 
 
 \begin{prop}
At any point $(t,z)\in \mathcal{C}^{\prime}$, the strategies that yield the minimum value of the functional $J$, (\ref{newJ}), are the optimal strategies for the problem (\ref{value.f}) at the point $(t,x)=(t, \mathcal{L}^{-1}_1(t,\cdot)(z)) \in \mathcal{C}$. 
\begin{proof}
Notice that at the terminal time $T$ there is the equality  
$$Z(T;t,z,\pi^1(\cdot), \pi^2(\cdot))=X(T;t,x,\pi^1(\cdot), \pi^2(\cdot)).$$
So,  we have the equality 
$$J(t,z,\pi^1(\cdot), \pi^2(\cdot))=\tilde{J}(t,x,\pi^1(\cdot), \pi^2(\cdot)),$$ 
which trivially  concludes  the proposition. 
\end{proof}
\end{prop}
By assuming some smoothness property, the strict decreasing property of the value function is proved in Subsection \ref{subclassical}. However, the decreasing property is obtained directly from the definitions. 
\begin{prop}
For any $t\in[0,T]$ and $S\leq z_1<z_2\leq F$, we have 
$$V(t,z_1)\geq V(t,z_2).$$ 
\begin{proof}
For any admissible strategy $(\pi^1, \pi^2)\in \Pi_{ad}(t,z_1)$, we have  
\begin{align}
X(T;t,z_1,\pi^1(\cdot),\pi^2(\cdot)) \leq X(T;t,z_2,\pi^1(\cdot),\pi^2(\cdot)).\label{decreasing}
\end{align}
Notice that two processes $X(\cdot;t,z_1,\pi^1(\cdot),\pi^2(\cdot))$ and  $X(\cdot;t,z_2,\pi^1(\cdot),\pi^2(\cdot))$ are continuous. So, if they are equal for some $s\in (t,T]$
then they are equal from that $s$ on. This observation implies that, by applying the strategy $(\pi^1, \pi^2)$ at the point $(t, z_2)$ the wealth process will remain upper than, or be equal to, the safety level $S$ which means that this strategy belongs to $\Pi_{ad}(t,z_2)$. Now, due to the definition of the loss function (\ref{lossfunc1}), the inequality (\ref{decreasing}) indicates a similar inequality for the corresponding action functional $J$, which concludes the claim. 
\end{proof}
\end{prop}

 
\subsection{Viscosity solution}
To prove that the value function (\ref{value1}) satisfy the HJB Equation (\ref{maineq}) in the viscosity sense, the continuity of the value function is required. To this end, we prove at first it is convex as a function of the space variable. 

\begin{prop}\label{convprop}
For any $t\in [0, T]$, the function $[S, F]\rightarrow \Real^+$, $x\rightarrow V(t,x)$ is strictly convex.  

\begin{proof}
The definitions of objective functionals $J, \tilde{J}$ imply the following equality for any $0\leq t\leq T$ and $S(t)\leq x\leq F(t)$, 
\begin{align}
\tilde{V}(t,x)=V(t, \mathcal{L}_1(t,x)). \label{V=V}
\end{align}
Moreover, the definition (\ref{diffeo}) shows that for a fixed $0\leq t\leq T$, the function $x\rightarrow \mathcal{L}_1(t,x)$ is linear. So, the convexity of the functions $x\rightarrow \tilde{V}(t,x)$ and   $x\rightarrow V(t,x)$ are equivalent. Here, we prove the convexity of the former function. 

For $\delta >0$, let $\pi^{1,\delta, x}(\cdot), \pi^{2,\delta, x}(\cdot)$ and $\pi^{1,\delta, y}(\cdot), \pi^{2,\delta, y}(\cdot)$ be $\delta$-optimal controls corresponding to the points $(t, x)$ and $(t, y)$, respectively, or 
$$\tilde{J}(t,x;\pi^{1,\delta, x}(\cdot), \pi^{2,\delta, x}(\cdot))\leq \tilde{V}(t,x)+\delta, \qquad  \tilde{J}(t,y;\pi^{1,\delta, y}(\cdot), \pi^{2,\delta, y}(\cdot))\leq \tilde{V}(t,y)+\delta.$$
Set $X^{\delta}_s:=X(s;t,x,\pi^{1,\delta, x}(\cdot), \pi^{2,\delta, x}(\cdot))$ and $Y^{\delta}_s:=X(s;t,y,\pi^{1,\delta, y}(\cdot), \pi^{2,\delta, y}(\cdot))$, for $t \leq s\leq T$. Then, for a fixed $\gamma\in [0, 1]$, setting $Z^{\delta}_s:=\gamma X^{\delta}_s+(1-\gamma)Y^{\delta}_s$ and $\pi^{2,\delta, z}_s:=\gamma \pi^{2,\delta, x}_s+(1-\gamma) \pi^{2,\delta, y}_s$, we get
\begin{align}
\gamma \tilde{V}(&t,x)+(1-\gamma) \tilde{V}(t,y)+\delta\nonumber\\
 &\geq \gamma \tilde{J}(t,x;\pi^{1,\delta, x}(\cdot), \pi^{2,\delta, x}(\cdot))+(1-\gamma) \tilde{J}(t,y;\pi^{1,\delta, y}(\cdot), \pi^{2,\delta, y}(\cdot))\nonumber\\
&=\gamma \mathbb{E}[\kappa \int^T_t \eta_s (C_1-\pi^{2,\delta, x}_s)^2ds+\eta_T \left(\frac{F-X^{\delta}_T}{a_{a_1}}\right)^2 ]\nonumber\\
 &\quad +(1-\gamma) \mathbb{E}[\kappa \int^T_t \eta_s (C_1-\pi^{2,\delta, y}_s)^2ds+\eta_T \left(\frac{F-Y^{\delta}_T}{a_{a_1}}\right)^2 ]\nonumber\\
 &\geq  \mathbb{E}[\kappa \int^T_t \eta_s (C_1-\pi^{2,\delta, z}_s)^2ds+ \eta_T \left(\frac{F-Z^{\delta}_T}{a_{a_1}}\right)^2 ],\label{dir}
\end{align}
where the last inequality is obtained from the convexity of $x\rightarrow (C_1-x)^2$ and $x\rightarrow (\frac{F-x}{a_{a_1}})^2$. 
Moreover, setting the control 
$$\pi^{1,\delta, z}_s:=\frac{1}{Z^{\delta}_s} \left( \gamma \pi^{1,\delta, x}_sX^{\delta}_s+(1-\gamma) \pi^{1,\delta, y}_sY^{\delta}_s \right ),  \qquad t \leq s\leq T,$$ 
we conclude from the dynamics  (\ref{wealtheq2})  that 
$$Z^{\delta}_s=X(s;t, \gamma x+(1-\gamma)y, \pi^{1,\delta, z}(\cdot), \pi^{2,\delta, z}(\cdot)).$$ 
Hence, the definition of the value function $\tilde{V}$ indicates 
\begin{align}
\tilde{V}(t,\gamma x +(1-\gamma) y)\leq \mathbb{E}[\kappa \int^T_t \eta_s (C_1-\pi^{2,\delta, z}_s)^2ds+ \eta_T \left(\frac{F-Z^{\delta}_T}{a_{a_1}}\right)^2 ].\label{rev}
\end{align}
Since $\delta$ is arbitrary in (\ref{dir}), we conclude from (\ref{dir}) and (\ref{rev}) the following inequality which indicates the convexity of $x\rightarrow \tilde{V}(t,x)$
\begin{align}
\tilde{V}(t,\gamma x +(1-\gamma) y)\leq \gamma \tilde{V}(t,x)+(1-\gamma) \tilde{V}(t,y).\label{finalineq}
\end{align}

Now, by contradiction suppose that $\tilde{V}$ is not strictly convex which means that (\ref{finalineq}) must be an equality. Hence, as $\delta \rightarrow 0$, the last inequality in (\ref{dir}) turns into an equality.  Then, from the strict convexity of $x\rightarrow (C_1-x)^2$ and $x\rightarrow (\frac{F-x}{a_{a_1}})^2$, we conclude that $\pi^{2,\delta, x}(\cdot)\rightarrow \pi^{2,\delta, y}(\cdot)\; a.s.$ and $X^{\delta}_T\rightarrow Y^{\delta}_T\; a.s.$, as $\delta \rightarrow 0$. Since the control variables $\pi^{2,\delta, x}, \pi^{2,\delta, y}$ are chosen from the bounded interval $[C_2, C_1]$, using the dominated convergence theorem and the dynamics (\ref{wealtheq2}), we conclude that, under the risk-neutral measure, the drift coefficient of the following It\^o process is $rM_s$,  
\begin{align}
M_s:=\lim_{\delta\rightarrow 0}(X^{\delta}_s-Y^{\delta}_s), \qquad s\geq t.\nonumber
\end{align}
 So, the discounted process $\tilde{M}_s=e^{-r(s-t)}M_s , s\geq t,$ is a martingale under the risk-neutral measure with initial value $\tilde{M}_t=x-y\neq 0 $. Therefore, under any equivalent measure $\tilde{\mathbb{P}}$ we have $\tilde{\mathbb{P}}\{M_T=0\}<1$ or equivalently 
$$\tilde{\mathbb{P}}\{\lim_{\delta\rightarrow 0}X^{\delta}_T\neq \lim_{\delta\rightarrow 0}Y^{\delta}_T\}>0,$$
which is a contradiction. 
\end{proof} 
\end{prop}

\begin{prop}\label{con}
The value function $V$ is continuous on the domain $[0, T]\times [S, F]$. 
\begin{proof}
the proof is in the Appendix.
\end{proof}
\end{prop}
The continuity of the value function supports the following theorem. 

\begin{thm}\label{vis}
The value function $V$ is the unique viscosity solution of the HJB Equation (\ref{maineq}) with boundary conditions (\ref{boundary})-(i)-(ii)-(iii) or (\ref{boundary})-(i)-(ii)-$(iii^{\prime})$.    
\begin{proof}
It can be checked easily that the coefficients of differential operator $\mathcal{A}$, and the running cost term $\kappa \eta_t(C_1-\pi^2)^2$ have continuous partial derivatives with respect to the variables $t$ and $z$,  for any fixed control variable. Furthermore, the values of control variables $\pi^1, \pi^2$ lie in the compact intervals. So, due to the continuity of value function, \cite[V. Theorem 3.1]{kn:FS} together with \cite[V. Corollary 3.1]{kn:FS} imply that $V$ is the viscosity solution of (\ref{maineq}). Moreover, \cite[V, Corollary 8.1]{kn:FS} indicates the uniqueness of viscosity solution of (\ref{maineq}) with boundary conditions (\ref{boundary})-(i)-(ii)-$(iii^{\prime})$. 

In addition, since the Dirichlet-type and Neumann-type boundary conditions in (\ref{boundary})-(i)-(ii)-(iii) are on unconnected parts of the boundary, on $z=S$ and $z=F$, respectively, we can apply  \cite[Theorem 3.1]{kn:B} on $z=F$, to conclude the comparison principle and then the uniqueness of viscosity solution.  
 \end{proof}
 \end{thm}
 
\subsection{Classical solution and the dual equation}\label{subclassical}
To apply the verification theorem, we must show that $V \in \mathcal{C}^{1,2}([0,T)\times(S,F);\Real)$. Our argument is similar to the one that has been employed in \cite{kn:DG, kn:FGG}. In fact,  the dual equation of the HJB Equation (\ref{maineq}) is considered. 

Let the viscosity solution of Equation (\ref{maineq}), $V$, belongs to $\mathcal{C}^{1,2}([0,T),(S,F);\Real)$ which implies that $V$ is a classical solution too. Then, due to strict convexity of V, which indicates 
\begin{align}
V_{zz}(t,z) > 0,\qquad \quad  \qquad \qquad (t,z)\in [0,T]\times(S,F),\label{conv}
\end{align}
 the investment control variable, that yields the  infimum in (\ref{maineq}), is given by 
\begin{align}
\pi^{1,*}=\frac{-(\mu-r)V_z(t,z)}{G(t)\sigma^2K(t,z)V_{zz}(t,z) }.\label{pi1}
\end{align}
Moreover, since the consumption control variable must be in the interval $[C_2, C_1]$, we get 
\begin{align}
  \pi^{2,*}=(\frac{G(t)V_z(t,z)}{2\kappa\eta_t}+C_1)\vee C_2. \label{pi2}
\end{align}
The formula (\ref{pi2}), divides the domain into two parts: 
\begin{align}
A=\{(t,z)\in [0,T)\times (S,F): \frac{G(t)V_z(t,z)}{2\kappa\eta_t}+C_1 > C_2  \},\label{setA}
\end{align}
\begin{align}
B=\{(t,z)\in [0,T)\times (S,F): \frac{G(t)V_z(t,z)}{2\kappa\eta_t}+C_1 \leq  C_2  \}. \label{setB}
\end{align}
Now, inserting the control variables (\ref{pi1}) and (\ref{pi2}) into (\ref{maineq}), V satisfies the following equations in classical sense 

\begin{align}
V_t(t,z)&+\left(H^{\prime}(t)+K(t,z)G^{\prime}(t)+rK(t,z)G(t)-C_1G(t) \right)V_z(t,z)\nonumber\\
&-\frac{1}{2}\frac{\beta^2V^2_z(t,z)}{V_{zz}(t,z)}-\frac{1}{4}\frac{G^2(t)V^2_z(t,z)}{\kappa\eta_t}=0, \qquad \quad (t,z)\in A, \label{eqA}
\end{align}

\begin{align}
V_t(t,z)&+\left(H^{\prime}(t)+K(t,z)G^{\prime}(t)+rK(t,z)G(t)-C_2G(t) \right)V_z(t,z)\nonumber\\
&-\frac{1}{2}\frac{\beta^2V^2_z(t,z)}{V_{zz}(t,z)}+\kappa \eta_t( C_1- C_2 )^2=0, \qquad \quad (t,z)\in B. \label{eqB}
\end{align}
It should be noted that the coefficients of the above two equations are the same on the common boundary  of $A$ and $B$.

The relation (\ref{conv}) together with (\ref{boundary})-(iii), imply   
\begin{align}
V_z(t,z)<0, \qquad \quad \qquad \qquad (t,z)\in [0,T]\times(S,F). \label{deriv}
\end{align}
In addition, we assume 
\begin{align}
\lim_{z\downarrow S}V_z(t,z)=-\infty, \qquad\qquad \qquad t\in [0, T). \label{VS}
\end{align}
Intuitively, this assumption means that the marginal loss when the wealth process approaches the safety level is very large, which seems to be reasonable. 

Now, we are ready to define the dual transformation. The relations (\ref{conv}), (\ref{deriv}), (\ref{boundary})-(iii) and (\ref{VS}) indicate that, for every $(t,y)\in [0,T)\times [0,+\infty)$, there is a unique minimizer $g(t,y)\in (S,F]$ of the function $[S,F]\rightarrow \Real^+, \; z\rightarrow V(t,z)+zy$. Moreover, it is characterized by the equation 
\begin{align} 
V_z(t,g(t,y))=-y, \qquad \qquad  (t,y) \in [0,T)\times [0,+\infty). \label{dual}
\end{align}
This characterization together with (\ref{VS}) imply 
\begin{align}
(i) g(t,y) \in (S,F), \qquad (ii) \lim_{y\rightarrow +\infty}g(t,y)=S, \quad  (t,y)\in [0,T)\times (0,+\infty). \label{insf}
\end{align}
Furthermore, from (\ref{conv}) and (\ref{dual}), we conclude that $g$ is differentiable in the space variable and  
\begin{align}
g_y(t,y) <0, \qquad \qquad \quad \qquad  (t,y)\in [0,T)\times (0,+\infty). \label{g_y}
\end{align}

Analogous to the subsets $A$ and $B$, defined in (\ref{setA}) and (\ref{setB}), the domain $[0,T)\times (0,+\infty)$ is divided to two subsets: 
\begin{align}
A^{\prime}=\{ (t,y)\in [0,T)\times (0,+\infty):y < \frac{2\kappa \eta_t}{G(t)}(C_1-C_2) \}, \nonumber
\end{align}
\begin{align}
B^{\prime}=\{ (t,y)\in [0,T)\times (0,+\infty):y \geq  \frac{2\kappa \eta_t}{G(t)}(C_1-C_2)\}. \nonumber
\end{align}
In the next proposition, we derive from the fully nonlinear Equations (\ref{eqA})-(\ref{eqB}) with the boundary conditions (\ref{boundary})-(i)-(ii)-(iii) the following equations for the function $g$ on $A^{\prime}$ and $B^{\prime}$ with the  boundary conditions (\ref{boundg}). 
\begin{align}
g_t(t,y)&+ (\beta^2 -K_1(t) G^{\prime}(t)-rK_1(t)G(t) ) y g_y(t,y)+\frac{1}{2}\beta^2y^2g_{yy} \nonumber\\
&-H^{\prime}(t)-K(t,g(t,y))G^{\prime}(t)-rK(t,g(t,y))G(t)\nonumber\\
&+C_1G(t)-\frac{1}{2} \frac{G^2(t)}{\kappa \eta_t } y =0, \qquad \qquad  \qquad\qquad \qquad (t,y)\in A^{\prime}, \label{eqgA} 
\end{align}

\begin{align}
g_t(t,y)&+ (\beta^2 -K_1(t) G^{\prime}(t)-rK_1(t)G(t) ) y g_y(t,y)+\frac{1}{2}\beta^2y^2g_{yy} \nonumber\\
&-H^{\prime}(t)-K(t,g(t,y))G^{\prime}(t)-rK(t,g(t,y))G(t)\nonumber\\
&+C_2G(t) =0, \qquad \qquad \qquad \qquad  \qquad \qquad \qquad \qquad (t,y)\in B^{\prime}. \label{eqgB}
\end{align}
It should be noted that the coefficients of the above two equations are the same on the common boundary of $A^{\prime}$ and $B^{\prime}$. More specifically, we have the equality $C_2G=C_1G-\frac{1}{2} \frac{G^2(t)}{\kappa \eta_t } y$ on this boundary. 

Considering the condition (\ref{boundary})-(iii), and finding the minimizer of the function $z\rightarrow V(T,z)+zy$ explicitly, we get the following boundary conditions: 
\begin{align}
\begin{cases}\label{boundg}
(i)\; g(t,0)=F, \qquad \qquad \qquad \qquad \qquad \quad  t\in [0,T], \\
(ii)\; g(T,y)=(F-\frac{a^2_{a_1}y}{2\eta_T})\vee S,\qquad \qquad y\in [0,+\infty ).
\end{cases}
\end{align}

Here we prove a proposition similar to \cite[Prop. 4.18]{kn:DG}. 

\begin{prop}
Suppose that the unique viscosity solution $V$ of (\ref{maineq}) with the boundary conditions (\ref{boundary})-(i)-(ii)-(iii) belongs to the class $\mathcal{C}^{1,3}([0,T)\times (S,F);\Real)$ and satisfies (\ref{VS}). Let $g$ be defined as above. Then $g$ is a classical solution of (\ref{eqgA})-(\ref{eqgB}) with the boundary conditions (\ref{boundg}). Moreover, $g$ satisfies (\ref{insf}) and (\ref{g_y}). 

Conversely, let $g\in \mathcal{C}([0, T]\times[0,+\infty );\Real)\cap \mathcal{C}^{1,2}([0,T)\times(0,+\infty);\Real) $ be a classical solution of (\ref{eqgA})-(\ref{eqgB})-(\ref{boundg}) and satisfies (\ref{insf}) and (\ref{g_y}). Furthermore, suppose 
\begin{align}
\lim_{y\rightarrow +\infty}y^2g_y(t,y)=0, \; uniformly\; in\; t\in [0,T), \label{yg_y}
\end{align}
\begin{align}
[g(t,\cdot)]^{-1} is \; integrable \; at \; S^+, \quad \qquad t\in [0,T),\label{intg}
\end{align}
and 
\begin{align}
\begin{cases}\label{boundh}
(i)\;h(t,z):=\int^T_t\kappa \eta_s(C_1- C_2)^2ds+\eta_T(\frac{F-S}{a_{a_1}})^2\\
\hspace{1.9cm} -\int^z_S[g(t,\cdot)]^{-1}(\zeta)d\zeta, \qquad \qquad  \quad  (t,z)\in[0,T)\times [S,F],\\
(ii)\;h(T,z):=\eta_T(\frac{F-z}{a_{a_1}})^2,\quad \qquad \qquad \qquad\qquad \qquad z\in[S,F].
\end{cases}
\end{align}
Then $h\in\mathcal{C}([0,T]\times[S,F])\cap \mathcal{C}^{0,1}([0,T)\times (S,F];\Real)\cap \mathcal{C}^{1,3}([0,T)\times (S,F);\Real)$. Furthermore, It is a classical solution of (\ref{maineq}) with boundary conditions (\ref{boundary})-(i)-(ii)-(iii) and satisfies  (\ref{VS}).

\begin{proof}
Let the unique viscosity solution $V$ of  (\ref{maineq}) with boundary conditions  (\ref{boundary})-(i)-(ii)-(iii) belongs to the class $\mathcal{C}^{1,3}([0,T)\times (S,F);\Real)$. Then, $V$ satisfies Equations (\ref{eqA})-(\ref{eqB}) in classical sense. Deriving these equations with respect to $z$ yield the following equations on $A$ and $B$, respectively. 

\begin{align}
V_{tz}&+(K_1G^{\prime}+rGK_1)V_{z}-\frac{\beta^2}{2}\frac{2V_zV^2_{zz}-V^2_zV_{zzz}}{V^2_{zz}}\nonumber\\
&+(KG^{\prime}+rGK+H^{\prime}-C_1G )V_{zz}-\frac{G^2V_zV_{zz}}{2\kappa\eta_t}=0, \label{V_tzA}
\end{align}
\begin{align}
V_{tz}&+(K_1G^{\prime}+rGK_1)V_{z}-\frac{\beta^2}{2}\frac{2V_zV^2_{zz}-V^2_zV_{zzz}}{V^2_{zz}}\nonumber\\
&+(KG^{\prime}+rGK+H^{\prime}-C_2G )V_{zz}=0. \label{V_tzB}
\end{align}
To shorten the above two formulas, the variables $t$ and $z$, inside the parentheses, have been eliminated. 

Since $V\in \mathcal{C}^{1,3}([0,T)\times (S,F);\Real)$, we have $g\in\mathcal{C}^{1,2}([0,T)\times(0, +\infty);\Real)$. Now, deriving Equation (\ref{dual}) w. r. t. $t$, w. r. t. $z$ and twice w. r. t. $z$, we get 
\begin{align}
&V_{tz}(t,g(t,y))+V_{zz}(t,g(t,y))g_t(t,y)=0,\label{1}\\
&V_{zz}(t,g(t,y))g_y(t,y)=-1,\label{2}\\
&V_{zzz}(t,g(t,y))g^2_y(t,y)+V_{zz}(t,g(t,y))g_{yy}(t,y)=0.\label{3}
\end{align} 
Due to the above relations, we  obtain Equations (\ref{eqgA})-(\ref{eqgB}) from Equations (\ref{V_tzA})-(\ref{V_tzB}). 
Furthermore, the boundary conditions (\ref{boundg}) and the properties (\ref{insf}) and (\ref{g_y}) have been demonstrated in the construction of the function $g$. \\

Conversely, let $g$ be given as in the statement. The relations (\ref{boundg}), (\ref{insf}) and (\ref{g_y}) clearly indicate that the function $[g(t,\cdot)]^{-1}$ is well-defined on $(S, F]$ for every $t\in [0,T)$. Furthermore, the definition (\ref{boundh}) indicates that 
$$h\in\mathcal{C}([0,T]\times[S,F] )\cap \mathcal{C}^{0,1}([0,T)\times(S,F];\Real)\cap\mathcal{C}^{1,3}([0,T)\times(S,F);\Real). $$
Moreover, the conditions (\ref{boundg}) and definitions (\ref{boundh}) imply (\ref{boundary})-(i)-(ii)-(iii).
Deriving the equation (\ref{boundh})-(i) with respect to $z$ and regarding (\ref{insf})-(ii), (\ref{VS}) is obtained with $h$ in place of $V$. 

Now, determining $h_z$ using the definition (\ref{boundh})-(i), we see that $h$ satisfies (\ref{dual}) with $h$ in place of $V$. Then, the argument of the first part shows that $h$ satisfies  (\ref{1})-(\ref{2})-(\ref{3}) with $h$ in place of $V$. Setting $z=g(t,y)$, applying backward the argument of the first part of the proof and due to the relation $g(t,(0,+\infty))=(S,F)$, we conclude that $h$ satisfies  Equations (\ref{V_tzA})-(\ref{V_tzB}) with $h$ in place of $V$.  Hence, by integrating these equations with respect to $z$, we get  
\begin{align}
h_t(t,z)&+\left(H^{\prime}(t)+K(t,z)G^{\prime}(t)+rK(t,z)G(t)-C_1G(t) \right)h_z(t,z)\nonumber\\
&-\frac{1}{2}\frac{\beta^2h^2_z(t,z)}{h_{zz}(t,z)}-\frac{1}{4}\frac{G^2(t)h^2_z(t,z)}{\kappa\eta_t}=C(t), \qquad \quad (t,z)\in A, \label{eqhA}
\end{align}

\begin{align}
h_t(t,z)&+\left(H^{\prime}(t)+K(t,z)G^{\prime}(t)+rK(t,z)G(t)-C_2G(t) \right)h_z(t,z)\nonumber\\
&-\frac{1}{2}\frac{\beta^2h^2_z(t,z)}{h_{zz}(t,z)}+\kappa \eta_t( C_1- C_2 )^2=C^{\prime}(t), \qquad \quad (t,z)\in B. \label{eqhB}
\end{align}
To conclude the claim, we must show that $C(t)=C^{\prime}(t)=0$, for any $t\in [0,T)$. Notice that the common boundary of the subsets $A$ and $B$ is a curve from $t=0$ to $t=T$. Since $h\in \mathcal{C}^{1,3}([0,T)\times(S,F);\Real)$, considering the above two equations on this curve, we conclude that $C\equiv C^{\prime}$. So, it is enough to prove $C^{\prime}\equiv 0$. 

Since $C^{\prime}$ is a time variable function, for any $z_0\in (S,F)$, we can write 
\begin{align}
C^{\prime}(t)= &h_t(t,z_0)-\frac{1}{2}\frac{\beta^2h^2_z(t,z_0)}{h_{zz}(t,z_0)}+\kappa \eta_t( C_1- C_2 )^2\nonumber\\
&+\left(H^{\prime}(t)+K(t,z_0)G^{\prime}(t)+rK(t,z_0)G(t)-C_2G(t) \right)h_z(t,z_0).\nonumber
\end{align}
Integrating both sides of the above equation over the time interval $[t,T]$ and regarding the definitions (\ref{boundh}), we have 
\begin{align}
\int^T_t C^{\prime}(s)ds=&h(T,z_0)- h(t,z_0)-\frac{\beta^2}{2}\int^T_t \frac{h^2_z(s,z_0)}{h_{zz}(s,z_0)}ds+\int^T_t \kappa \eta_s( C_1- C_2 )^2ds \nonumber\\
&+\int^T_tD(s,z_0) h_z(s,z_0)ds \nonumber\\
=&\eta_T(\frac{F-z_0}{a^2_{a_1}} )^2-\eta_T(\frac{F-S}{a^2_{a_1}})^2+\int^{z_0}_S [g(t,\cdot)]^{-1}(\zeta)d\zeta \nonumber\\
&-\frac{\beta^2}{2}\int^T_t \frac{h^2_z(s,z_0)}{h_{zz}(s,z_0)}ds+\int^T_tD(s,z_0)h_z(s,z_0)ds,\label{Ct}
\end{align}
in which 
\begin{align}
D(s,z)=H^{\prime}(s)+K(s,z)G^{\prime}(s)+rK(s,z)G(s)-C_2G(s).\label{Dfunc}
\end{align}
Then, taking $z_0\downarrow S$ in the above formula, we get 
\begin{align}
\int^T_t C^{\prime}(s)ds=\lim_{z_0\downarrow S} -\frac{\beta^2}{2}\int^T_t \frac{h^2_z(s,z_0)}{h_{zz}(s,z_0)}ds+\lim_{z_0\downarrow S}\int^T_tD(s,z_0)h_z(s,z_0)ds.\label{convC}
\end{align}
For a given $(s,y_0)\in [0,T)\times (0,+\infty)$, set $z_0(s)=g(s,y_0)$. Due to (\ref{2}) and (\ref{dual}), we have 
$$\frac{h^2_z(s,z_0(s))}{h_{zz}(s,z_0(s))}=-y^2_0g_y(s,y_0). $$ 
The relation (\ref{insf})-(ii) implies that $z_0(s) \downarrow S$, when $y_0\rightarrow +\infty$. Moreover, this convergence is uniform with respect to $s \in [0,T)$; (see \cite[Prop. 4.18]{kn:DG}). 
So, we obtain the following convergence due to the condition (\ref{yg_y})
\begin{align}
\lim_{z_0\downarrow S}\int^T_t\frac{h^2_z(s,z_0)}{h_{zz}(s,z_0)}ds=-\lim_{y_0\rightarrow +\infty}\int^T_t y^2_0g_y(s,y_0)ds=0. \label{C1}
\end{align}

Now, taking $z$ equal to $S$ in (\ref{Dfunc}) and regarding the formulas in (\ref{Gfunc}), (\ref{zproc}) and (\ref{K}), we get 
\begin{align}
D(s,S)=&\frac{C_2F-C_1S}{F(s)-S(s)}e^{-r(T-s)}\nonumber\\
&+\frac{(C_2F-C_1S)(C_2-C_1+r(F-S))}{(F(s)-S(s))^2}\frac{e^{-r(T-s)}-e^{-2r(T-s)}}{r}\nonumber\\
&-\frac{S(C_2-C_1+r(F-S))e^{-r(T-s)} }{F(s)-S(s)}\nonumber\\
&+\frac{[C_2F-C_1S](e^{-r(T-s)}-1)}{r(F-S)}\frac{(F-S)(C_2-C_1+r(F-S))e^{-r(T-s)} }{(F(s)-S(s))^2}\nonumber\\
&+\frac{r[F(s)-S(s)]S}{(F-S)}\frac{F-S}{F(s)-S(s)}\nonumber\\
&-\frac{[C_2F-C_1S]e^{-r(T-s)}}{(F-S)}\frac{F-S}{F(s)-S(s)}\nonumber\\
&+\frac{[C_2F-C_1S]}{F-S}\frac{F-S}{F(s)-S(s)}\nonumber\\
&-\frac{C_2(F-S)}{F(s)-S(s)}:=e_1+e_2+e_3+e_4+e_5+e_6+e_7+e_8,\nonumber
\end{align}
where $H^{\prime}(s)=e_1+e_2$, $K(s,S)G^{\prime}(s)=e_3+e_4$ and $rK(s,S)G(s)=e_5+e_6+e_7$. Clearly $e_1+e_6=0$ and $e_2+e_4=0$. Moreover, regarding the identity 
\begin{align}
F(s)-S(s)=\frac{C_1-C_2}{r}+\frac{1}{r}(C_2-C_1+r(F-S))e^{-r(T-s)},\label{F-S}
\end{align}
 we get $e_3=-rS+\frac{S(C_1-C_2)}{F(s)-S(s)}$, which yields
 $$e_3+e_5+e_7+e_8=-rS+\frac{S(C_1-C_2)}{F(s)-S(s)}+rS+\frac{C_2F-C_1S}{F(s)-S(s)}-\frac{C_2(F-S)}{F(s)-S(s)} = 0.$$ 
 Therefore, we have the identity $D(\cdot,S)\equiv 0$. Since the function $D$ is continuous, this identity implies 
\begin{align}
\lim_{z\downarrow S}D(s,z)=0, \qquad \qquad s\in [0,T].\nonumber
\end{align}
Furthermore, regarding the definition of the function $K$, (\ref{K}),  $D$ is a linear function of the variable $z$. Hence, due to the boundedness of the functions in (\ref{Dfunc}), the above convergence is  uniform with respect to $s$ on the compact interval $[0,T]$. 

On the other hand, the integrability of $[g(s,\cdot)]^{-1}$ at $S^+$, Condition (\ref{intg}), indicates that the convergence $\lim_{z\downarrow S}h_z(s,z)=-\infty$ is sufficiently slow that  its multiplication by a linear function of $z$, the integrand $D(s,z_0)h_z(s,z_0)$, tends to zero, uniformly over $[0,T]$, as $z_0\downarrow S$. This implies  the convergence 
\begin{align}
\lim_{z_0\downarrow S}\int^T_tD(s,z_0)h_z(s,z_0)ds=0.\label{C2}
\end{align}

Since $C^{\prime}$ is a continuous function and $t\in [0,T]$ is arbitrary in (\ref{convC}), the convergences (\ref{C1}) and (\ref{C2}) yield the identity $C^{\prime}\equiv 0$.
\end{proof}

\end{prop}

\begin{cor}
Since the function $h$ given in the above proposition is a classical solution of (\ref{maineq}) with conditions (\ref{boundary})-(i)-(ii)-(iii), it is as a viscosity solution too. Therefore, due to the uniqueness of viscosity solution, it is equal to the value function $V$. 
\end{cor}

Due to the above proposition, to prove the required regularity of the value function $V$, we must prove that there is a function $g$ that satisfies the assumptions of such proposition. This is the claim of the following theorem. 

\begin{thm}
There exists a unique $g\in\mathcal{C}([0,T]\times [0,+\infty);\Real)\cap\mathcal{C}^{1,2}([0,T)\times(0,+\infty);\Real)$ that is the classical solution of (\ref{eqgA})-(\ref{eqgB})-(\ref{boundg}) and  satisfies (\ref{insf}), (\ref{g_y}), (\ref{yg_y}) and (\ref{intg}). 

\begin{proof}
We write the Equations (\ref{eqgA})-(\ref{eqgB}) in a unified form which is a linear parabolic equation over the domain $[0,T)\times (0,+\infty)$, 
\begin{align}
g_t(t,y)+\mu(t,y)g_y(t,y)+\frac{1}{2}\beta^2 y^2g_{yy}(t,y)-q(t)g(t,y)+f(t,y)=0,\label{parabolic}
\end{align}
with the boundary conditions  
\begin{align}
\begin{cases}\label{parabound}
(i)\; g(t,0)=F, \qquad \qquad \qquad \qquad \quad \qquad   t\in [0,T],\\
(ii)\; g(T,y)=\Psi(y), \qquad \qquad \qquad \qquad \quad y\in [0,+\infty ),
\end{cases}
\end{align}
in which 
$$\mu(t,y)= (\beta^2 -K_1(t) G^{\prime}(t)-rK_1(t)G(t) ) y, $$
$$q(t)=K_1(t)G^{\prime}(t)+rK_1(t)G(t), $$
$$f(t,y)=-H^{\prime}(t)+(G^{\prime}(t)+rG(t))K_2(t)+G(t)(C_1- \frac{G(t)}{2\kappa \eta_t}y)\vee C_2,$$ 
$$\Psi(y)=(F-\frac{a^2_{a_1} y}{2\eta_T})\vee S.$$

\emph{Existence of viscosity solution}. At first we show that the above equation has a unique bounded viscosity solution. 

In the first step, we find the bounded sub and supersolution of Equation (\ref{parabolic}). See \cite[Section 2]{kn:CIL}, for the definition of these  types of solutions. 

Since the coefficients $\mu/y$, $q$ and $f$ are bounded functions, by taking their upper and lower bounds, we establish sub and supersolution. 

The explicit formula (\ref{qfunc}) indicates that $q$ is a nonnegative function which means that $ \beta^2y$ is an upper bound for $\mu$. Moreover, the explicit formula (\ref{f_explicit}) and the fact that $f_{\infty}\leq f$, which is clear from their definitions, indicate that $f$ is nonnegative. So, setting $q_1=\sup_{0\leq t\leq T}q(t)$, the solution of the following equation with conditions (\ref{parabound}) can be regarded as a subsolution to the Eq. (\ref{parabolic})   
\begin{align}
\bar{g}_t(t,y)+\beta^{2} y\bar{g}_y(t,y)+\frac{1}{2}\beta^2 y^2\bar{g}_{yy}(t,y)-q_1\bar{g}(t,y)=0, \quad (t,y)\in [0,T)\times (0,+\infty).\nonumber
\end{align}
This equation with the given conditions has a classical solution which can be represented as 
$$\bar{g}(t,y)=\frac{a^2_{a_1}}{2\eta_T}e^{(\beta^{2}-q_1)(T-t)} \bar{p}_{put}(t,y)+S \in [S,F],$$
 where $\bar{p}_{put}$ is the price of a European put option with strike price $\frac{2\eta_T}{a^2_{a_1}}(F-S)$ in a Black-Scholes market where the volatility of the risky asset equals to $\beta$ and the spot rate of riskless asset is $\beta^{2}$. 

To find a supersolution, setting $q_2=\beta^2-q_1$, we consider the following equation with the same conditions as in (\ref{parabound}). 
\begin{align}
\hat{g}_t(t,y)+q_2 y\hat{g}_y(t,y)+\frac{1}{2}\beta^2 y^2\hat{g}_{yy}(t,y)+f_0=0,  \quad (t,y)\in [0,T)\times (0,+\infty), \nonumber
\end{align}
in which $f_0= \sup_{t,y}f(t,y)$. The explicit solution of this equation can be written as 
\begin{align}
\hat{g}(t,y)=&\frac{a^2_{a_1}}{2\eta_T}e^{q_2(T-t)}\hat{p}_{put}(t,y)+S+f_0(T-t)\nonumber\\    
=&(F-S)\Phi(k(t,y))-\frac{a^2_{a_1}}{2\eta_T}e^{q_2(T-t)}y\Phi\left(k(t,y)-\beta \sqrt{T-t}\right )\nonumber\\
&+S+f_0(T-t), \label{closedform}
\end{align}
where 
$$k(t,y)=\frac{-\log(\frac{a^2_{a_1}y}{2\eta_T(F-S)})-\frac{\beta^2}{2}(T-t) }{\beta \sqrt{T-t}},$$ 
and $\Phi$ is the cumulative distribution function of a standard normal random variable. Actually, $\hat{p}_{put}$ is the price of a European put option with strike price $\frac{2\eta_T}{a^2_{a_1}}(F-S)$ in a Black-Scholes market where the volatility of the risky asset equals to $\beta$, and the spot rate of riskless  asset is $q_2$. Furthermore, we easily get $\hat{g}(t,y) \in [S, F+f_0T]$. 
 
Now, we use \cite[Theorem 8.2]{kn:CIL}, to conclude the comparison principle for the sub and supersolution in the class of bounded upper and lower semicontinuous functions. Although, in such a theorem a bounded domain is considered, the proof can be modified directly for the domain $(0,+\infty)$, as it is done in \cite[Theorem 9.1]{kn:FS} for the real line $(-\infty, +\infty)$. 

Due to the existence of bounded sub and supersolution and the comparison principle, and by applying Perron's method \cite[Theorem 4.1]{kn:CIL} we can conclude the existence of a unique bounded viscosity solution $g_{visc}$ such that $\bar{g}\leq g_{visc}\leq \hat{g}$ over $[0,T)\times [0, +\infty)$. \\

\emph{$\mathcal{C}^{1,2}$ regularity of the solution}. For any $0<a<b<+\infty$, consider the Equation (\ref{parabolic}) over $[0,T)\times (a,b)$ with the boundary conditions that are set by the viscosity solution 
\begin{align}    
&g_t(t,y)+\mu(t,y)g_y(t,y)+\frac{1}{2}\beta^2y^2g_{yy}(t,y)-q(t)g(t,y)+f(t,y)=0,\nonumber\\
&g(t,a)=g_{visc}(t,a), \qquad \quad g(t,b)=g_{visc}(t,b), \qquad \quad\; t\in [0,T), \nonumber\\
&g(T,y)=g_{visc}(T,y),\qquad \qquad \qquad \qquad \qquad \qquad \qquad y\in[0, +\infty).\nonumber
\end{align}
It is clear that the above equation has a unique viscosity solution which must be equal to $g_{visc}$. On the other hand, the equation on the given domain has uniformly parabolic property, which implies the existence of a classical solution that belongs to $\mathcal{C}^{1,2}([0,T)\times (a,b);\Real)$. Since the classical solution is also a viscosity solution, it must coincide with $g_{visc}$. 

Now, from the arbitrariness of $a$ and $b$ we conclude that $g_{visc}\in \mathcal{C}^{1,2}([0,T)\times (0,+\infty);\Real)$. Therefore, $g_{visc}$ is the unique classical solution of (\ref{parabolic}) with boundary conditions (\ref{parabound}). \\

\emph{Feynman-Kac formula and properties of solution}. After proving that Equation (\ref{parabolic}) with conditions (\ref{parabound}) has a classical solution, we can find, using the Feynman-Kac formula, a probabilistic representation of the solution, which reveals some of its properties 
\begin{align}
g(t,y)=\mathbb{E}^{\mathbb{Q}} [\int^T_t e^{-\int^s_t q(\tau) d\tau} f(s,Y^y_s)ds + e^{-\int^T_t q(\tau)d\tau}\Psi(T,Y^y_T) ],\label{Kac}
\end{align}
where the expectation is under the probability measure $\mathbb{Q}$ such that $Y^y$ is an It\^o process driven by the equation 
\begin{align}
\begin{cases}\nonumber
dY^y_s=\mu(s,Y^y_s)ds+\frac{1}{2}\beta^2(Y^y_s)^2dW^{\mathbb{Q}}_s,\qquad \qquad s > t, \\
Y^y_t=y,
\end{cases}
\end{align}
with $W^{\mathbb{Q}}$ being a Brownian motion under $\mathbb{Q}$.  

Since $f$ and $\Psi$ are decreasing functions of $y$, we can conclude from the Feynman-Kac formula that $g$ is strictly decreasing with respect to $y$; the relation (\ref{g_y}). Hence, the relation (\ref{parabound})(i) implies $g\leq F$ over $[0,T]\times [0,+\infty)$. 

Now, we aim to determine the asymptotic behavior of $g(t,y)$ when $y\rightarrow +\infty$. 
Considering the definition of $f$, its restriction to the subset $B^{\prime}$ is a function of $t$ and is denoted by $f_{\infty}(t)$
\begin{align}
f_{\infty}(t):=-H^{\prime}(t)+(G^{\prime}(t)+rG(t))K_2(t)+G(t) C_2. \label{dfunc}
\end{align} 
Moreover, it is clear that as $y\rightarrow +\infty$, $\mathbb{P}\{\exists s\in [t,T]: (s,Y^y_s)\in A^{\prime} \}\rightarrow 0$, for any $0\leq t\leq T$. 
On the other hand, we have $\lim_{y\rightarrow +\infty}\mathbb{P}(\Psi(T,Y^y_T)=S)=1$. Therefore, as $y\rightarrow +\infty$, $g(t,y)$ converges to the integral 
\begin{align}
g_{\infty}(t):=\mathbb{E}^{\mathbb{Q}}[\int^T_t e^{-\int^s_t q(\tau)d\tau}f_{\infty}(s)ds+e^{-\int^T_t q(\tau)d\tau} S ]\label{feynman}.
\end{align}  
To evaluate the above integral, the functions $q$ and $f_{\infty}$ must be written explicitly. 

From the formula of $q$, we get 
\begin{align}
q(t)=&K_1(t)G^{\prime}(t)+rK_1(t)G(t)\nonumber\\
=&\frac{F(t)-S(t)}{F-S}\frac{-(F-S)(C_2-C_1+r(F-S))e^{-r(T-t)}}{(F(t)-S(t))^2}+r\nonumber\\
=&\frac{C_1-C_2}{F(t)-S(t)},\label{qfunc}
\end{align}
where, in the last equality the relation (\ref{F-S}) is applied. 
Moreover, regarding the formula of $f_{\infty}$, we have 
\begin{align}
f_{\infty}(t)=&-\frac{C_2F-C_1S}{F(t)-S(t)}e^{-r(T-t)}\nonumber\\
&-\frac{(C_2F-C_1S)(C_2-C_1+r(F-S))}{(F(t)-S(t))^2}\frac{e^{-r(T-t)}-e^{-2r(T-t)}}{r}\nonumber\\
&+\frac{-(F-S)(C_2-C_1+r(F-S))e^{-r(T-t)}}{(F(t)-S(t))^2} \frac{[C_2F-C_1S](e^{-r(T-t)}-1)}{r(F-S)}\nonumber\\
&+r\frac{F-S}{F(t)-S(t)} \frac{[C_2F-C_1S](e^{-r(T-t)}-1)}{r(F-S)}\nonumber\\
&+\frac{C_2(F-S)}{F(t)-S(t)}=b_1+b_2+b_3+b_4+b_5.\nonumber
\end{align}
Clearly $b_2+b_3=0$. Therefore, a simple manipulation shows that 
\begin{align}
f_{\infty}(t)=\frac{S(C_1-C_2)}{F(t)-S(t)}=Sq(t). \label{f_explicit}
\end{align}
Hence, (\ref{feynman}) turns into 
\begin{align}
g_{\infty}(t)=S\mathbb{E}^{\mathbb{Q}}[\int^T_t e^{-\int^s_t q(\tau)d\tau}q(s)ds+e^{-\int^T_t q(\tau)d\tau} ].\label{gq}
\end{align}
In view of the indefinite integral calculation $\int e^{-\int^s_t q(\tau)d\tau}q(s)ds=-e^{-\int^s_t q(\tau)d\tau}$, we obtain  
$$\int^T_t e^{-\int^s_t q(\tau)d\tau}q(s)ds=-e^{-\int^s_t q(\tau)d\tau}]^T_t=-e^{-\int^T_t q(\tau)d\tau} +1.$$
Therefore, we get the following identity, which indicates (\ref{insf})-(ii), 
\begin{align}
g_{\infty}(t)=S, \qquad \qquad t\in [0,T].\nonumber
\end{align}

Now, we take into account the actual difference $g(t,y)-S$, when $y$ is very large. Due to the above identity and (\ref{Kac}), we can write 
\begin{align}
g(t,y)-S=&\mathbb{E}^{\mathbb{Q}}[\int^T_t e^{-\int^s_t q(\tau)d\tau} (f(s,Y^y_s)-f_{\infty}(s))ds]\nonumber\\
&+\mathbb{E}^{\mathbb{Q}}[e^{-\int^T_t q(\tau)d\tau}(\Psi(T,Y^y_T)- S)] \nonumber\\
=&d_1(t,y)+d_2(t,y).\nonumber
\end{align}
Note that $f(s,Y^y_s)=f_{\infty}(s)$ for $(s,Y^y_s)\in B^{\prime}$ and $\Psi(T,Y^y_T)=S$ when $Y^y_T\geq \frac{2\eta_T}{a^2_{a_1}}(F-S)$. So, by  rough estimates of $\mathbb{P}\{\exists s\in [t,T] :(s,Y^y_s)\in A^{\prime}\}$, which is an exit probability on a finite interval, and $\mathbb{P}\{ Y^y_T < \frac{2\eta_T}{a^2_{a_1}}(F-S)\}$, which is obtained from the log-normal distribution, we obtain the following upper bounds, for large enough $y$, 
 $$d_1(t,y)< \frac{c_0}{y^2}, \qquad  d_2(t,y)< \frac{c_1}{y^2},$$
 in which $c_0, c_1$ are positive constants. Then, for a constant $c_2$, that is large enough and greater than $c_0+c_1$, we get the estimate   
\begin{align}
\abs{g(t,y)-S}<\frac{c_2}{y^2}, \qquad \quad  (t,y)\in [0,T)\times (0,\infty), \label{g-s}
\end{align}
which implies the integrability condition (\ref{intg}). 
  
The coefficients of Equation (\ref{parabolic}) satisfy the conditions of \cite[Theorems 3.1 and 6.1]{kn:JT} which implies that our linear parabolic equation, with convex terminal condition $\Psi$,  preserves the convexity. Therefore, $g$ is convex in the variable $y$.  

From the convexity of $g$ and (\ref{g_y}), we achieve  
 \begin{align}
 y(g(t,2y)-g(t,y) )\leq y^2g_y(t,2y)<0, \qquad (t,y) \in [0,T)\times (0,+\infty). \label{g2y}
\end{align}
The estimate (\ref{g-s}) indicates $g(t,y)-g(t,2y) \leq \frac{c_2}{y^2}$ or $g(t,2y)-g(t,y) \geq -\frac{c_2}{y^2}$. So, regarding (\ref{g2y}), we have 
$$-\frac{8c_2}{y}\leq y^2g_y(t,y)<0, $$
 which yields the relation (\ref{yg_y}). 
\end{proof}
\end{thm}

\subsection{Verification theorem} 
After proving that the value function is the only classical solution of the Equations (\ref{eqA})-(\ref{eqB}), we can state the verification theorem which concerns the classical solution of (\ref{maineq}) and gives a way of testing whether the given admissible strategies are optimal. To see a standard proof of the theorem, we refer the reader to \cite[Ch. IV, Theorem 3.1]{kn:FS}. 

\begin{thm}{\textbf{(verification theorem)}}\\
(a) Let $h\in \mathcal{C}([0,T]\times[S,F] )\cap \mathcal{C}^{1,2}([0,T)\times(S,F);\Real)$ be a classical solution to (\ref{maineq}) with boundary conditions (\ref{boundary})-(i)-(ii)-($iii^{\prime}$). Then, for any initial data $(t,z)\in\mathcal{C}^{\prime}$ and admissible strategy $\pi=(\pi^1(\cdot), \pi^2(\cdot)) \in\Pi_{ad}(t,z)$, we have 
$$h(t,z)\leq J(t,z;\pi^1(\cdot), \pi^2(\cdot)). $$
(b) If there exists $\pi^*= (\pi^{1,*}(\cdot), \pi^{2,*}(\cdot)) \in \Pi_{ad}(t,z)$ such that for any $s\geq t$
\begin{align}
(\pi^{1,*}_s, \pi^{2,*}_s)\in {arg}\; {min} \{ \mathcal{A}h(s,Z^*_s)+\kappa \eta_s(C_1-\pi^{2,*}_s)^2 \},  \nonumber
\end{align}
in which the  operator $\mathcal{A}$ is defined in (\ref{diffA}) and $Z^*_s$ is the solution to (\ref{wealtheq2}) corresponding to the strategy  $\pi^{*}$, with $Z^*_t=z$, then $\pi^*$ is the optimal strategy and $h$ is equal to the value function, 
$$h(t,z)=V(t,z)=J(t,z;\pi^{1,*}(\cdot), \pi^{2,*}(\cdot)).$$
\end{thm}

At any point $(t,z)\in [0,T)\times [S,F]$, we establish the closed loop equation to prove the existence and uniqueness of the optimal strategies. At first, due to the formulas (\ref{pi1}) and (\ref{pi2}) and using (\ref{dual}) and (\ref{2}), we get the following formulas for feedback maps in terms of the function $g$   
\begin{align}
 P^1(t,z)=\begin{cases} -\frac{(\mu-r)[g(t,\cdot)]^{-1}(z)g_y(t,[g(t,\cdot)]^{-1}(z)) }{G(t)\sigma^2 K(t,z)}, &  (t,z)\in [0,T)\times (S,F), \\ 0, & (t,z)\in [0,T)\times \{S,F\}, \end{cases} \label{pi1g}
\end{align}
\begin{align}
P^2(t,z)=\begin{cases} (-\frac{G(t)[g(t,\cdot)]^{-1}(z)}{2\kappa\eta_t}+C_1)\vee C_2, & (t,z)\in [0,T)\times (S,F), \\ C_1, & (t,z)\in [0,T)\times \{F\}, \\C_2, & (t,z)\in [0,T)\times \{S\}. \end{cases}\label{pi2g}
\end{align}
The relation (\ref{yg_y}) implies that $P^1$ and $P^2$ are bounded and continuous functions on $[0,T)\times [S,F]$. 

For some $(t,z)\in [0,T)\times (S,F)$, let $y^*=[g(t,\cdot)]^{-1}(z)$ and suppose that $Y^*(\cdot;t,y^*)$ is the solution to 
\begin{align}
\begin{cases} dY^*_s=-(K_1(s)G^{\prime}(s)+rK_1(s)G(s))Y^*_s ds-\beta Y^*_s dW_s, &\quad  t < s< T, \\ Y^*_t=y^*. \end{cases}\label{Y^*}
\end{align}
Now, consider the process 
\begin{align}
Z^*(s;t,z)=g(s,Y^*(s;t,y^*)), \qquad \qquad  s\in [t,T].\label{Z^*}
\end{align}
Due to the definition of $Y^*$ and properties of $g$, we have 
\begin{align}
Z^*(s;t,z)\in (S,F),\qquad\qquad  \qquad \forall s\in [t,T].\label{ZinSF}
\end{align}

\begin{thm}{\textbf{(closed loop equation)}}
 For any $(t,z)\in [0,T)\times (S,F)$, $Z^*(\cdot;t,z)$ solves the following closed loop equation, associated with the feedback maps $P^1$ and $P^2$, 
\begin{align}
\begin{cases} 
dZ^*_s=\{ K(s,Z^*_s) G^{\prime}(s)+G(s) [(P^1(s,Z^*_s)(\mu-r)+r)K(s,Z^*_s)-P^2(s,Z^*_s) ]\nonumber\\ 
\qquad \quad\; +H^{\prime}(s) \} ds+G(s)\sigma P^1(s,Z^*_s) K(s,Z^*_s) dW_s, \qquad\qquad t < s< T, \nonumber\\
Z^*_t=z.                             
\end{cases} 
\end{align}

\begin{proof}
Regarding the definition (\ref{Z^*}), the dynamics (\ref{Y^*}) and the Equations (\ref{eqgA}), (\ref{eqgB}), and applying It\^o's formula, the following dynamics is obtained  
\begin{align}
dZ^*_s=&g_s(s,Y^*_s)ds-(K_1(s)G^{\prime}(s)+rK_1(s)G(s))Y^*_sg_y(s,Y^*_s)ds\nonumber\\
&+\frac{1}{2}\beta^2(Y^*_s)^2g_{yy}(s,Y^*_s)ds-\beta Y^*_sg_y(s,Y^*_s)dW_s\nonumber\\
=&(-\beta^2Y^*_sg_y(s,Y^*_s)+H^{\prime}(s)+K(s,Z^*_s)G^{\prime}(s)+rK(s,Z^*_s)G(s))ds\nonumber\\
&-G(s)((-\frac{1}{2} \frac{G(s)}{\kappa \eta_s }Y^*_s+C_1)\vee C_2)ds-\beta Y^*_sg_y(s,Y^*_s)dW_s.\nonumber
\end{align}
The definition (\ref{Z^*}) indicates $Y^*_s=[g(s,\cdot)]^{-1}(Z^*_s)$. So, regarding the formulas (\ref{pi1g}), (\ref{pi2g}), a simple manipulation yields the closed loop equation.  

Moreover, notice that the closed loop equation coincides with Equation (\ref{zproc2}), if the feedback maps be replaced by $(\pi^1, \pi^2)$. 
\end{proof}
\end{thm}
\begin{rem}
Due to formulas (\ref{pi1g}) and (\ref{pi2g}), the closed loop equation admits the solutions $Z^*(\cdot)\equiv F$ and $Z^*(\cdot)\equiv S$ corresponding to the initial values $z=F$ and $z=S$,  respectively. 
\end{rem}

\begin{cor} 
The boundedness of the maps $P^1$ and $P^2$, the relation (\ref{ZinSF}) and the above remark imply that, for any $(t,z)\in [0,T]\times [S,F]$, the following strategy is admissible. Moreover, the optimality is concluded from the verification and the closed loop equation theorems.
\begin{align}
\pi^1_s=\begin{cases} P^1(s,Z^*(s;t,z)), &  s\in[t , T), \\ 0, & s=T, \end{cases}\label{p1optimal}
\end{align}
\begin{align}
\pi^2_s=\begin{cases} P^2(s,Z^*(s;t,z)), &  s\in[t , T), \\ 0, & s=T. \end{cases}\label{p2optimal}
\end{align}
\end{cor}
The uniqueness of the optimal strategy is proved in the following. 
\begin{prop}
At any point $(t,z)\in [0,T]\times [S,F]$, the given above strategy, $\pi=(\pi^1, \pi^2)$, is the only optimal strategy. 
\begin{proof}
By contradiction let $\pi^{\prime}=(\pi^{1^{\prime}}, \pi^{2^{\prime}})$ be another optimal strategy at the point $(t,z)$. Similar to the argument used in Prop. (\ref{convprop}), defining $X^*_s:=\frac{1}{2}(X_s+X^{\prime}_s)$, in which $X_s=X(s;t,z,\pi^1(\cdot),\pi^2(\cdot))$, $X^{\prime}_s=X(s;t,z,\pi^{1^{\prime}}(\cdot),\pi^{2^{\prime}}(\cdot))$, and  
$$\pi^{1,*}_s:=\frac{1}{2X^*_s}(\pi^1_sX_s+\pi^{1^{\prime}}_sX^{\prime}_s),$$
$$\pi^{2,*}_s:=\frac{1}{2}(\pi^2_s+\pi^{2^{\prime}}_s),$$
we have $X^*_s=X(s;t,z,\pi^{1,*}(\cdot),\pi^{2,*}(\cdot))$. 

Now, due to the strict convexity of the functions $x\rightarrow (C_1-x)^2$ and $x\rightarrow (\frac{F-x}{a_{a_1}})^2$, the following strict inequality for the action functionals is concluded, which contradicts the optimality of the strategies $\pi$ and $\pi^{\prime}$, 
\begin{align}
\frac{1}{2}& [ J(t,z;\pi^1(\cdot), \pi^2(\cdot))+J(t,z;\pi^{1^{\prime}}(\cdot), \pi^{2^{\prime}}(\cdot)) ]\nonumber\\
&=\; \frac{1}{2} \mathbb{E}[\kappa \int^T_t \eta_s (C_1-\pi^{2}_s)^2 ds+\eta_T (\frac{F-X_T}{a_{a_1}})^2 ]\nonumber\\
 &\;\;\; \;\;+\frac{1}{2} \mathbb{E}[\kappa \int^T_t \eta_s (C_1-\pi^{2^{\prime}}_s)^2 ds+\eta_T (\frac{F-X^{\prime}_T}{a_{a_1}})^2 ]\nonumber\\
 &>  \mathbb{E}[\kappa \int^T_t \eta_s (C_1-\pi^{2,*}_s)^2ds+ \eta_T (\frac{F-X^*_T}{a_{a_1}})^2 ]=J(t,z;\pi^{1,*}(\cdot), \pi^{2,*}(\cdot)).\nonumber
\end{align}
\end{proof}
\end{prop} 
Due to (\ref{p1optimal}) and (\ref{p2optimal}), the optimal strategy is obtained by applying the feedback maps on the solution of the closed loop equation. So, the uniqueness of the optimal strategy yields the uniqueness of solution of the closed loop equation; see \cite[Remark 5.4]{kn:DG} for a rigorous proof. 


\section{\textbf{Numerical Algorithm}}
In Section 4, the Neumann boundary condition (\ref{boundary})-(iii) is employed to show the regularity of the value function $V$. However, due to Theorem \ref{vis}, the value function $V$ is the unique viscosity solution of (\ref{maineq}) with conditions (\ref{boundary})-(i)-(ii)-($iii^{\prime}$). Therefore, to get the numerical approximation of the value  function, we can use the Dirichlet condition (\ref{boundary})-($iii^{\prime}$) instead of the Neumann condition (\ref{boundary})-(iii). 

\subsection{Finite difference method} 
The finite difference method is applied to discretize Equation (\ref{maineq}). The time horizon $[0, T]$ is divided to $M=T\times 52$ subintervals of equal length $\Delta t=\frac{1}{52}$, the length of one week in a year. Moreover, the space interval $[S, F]$ is discretized as $S=z_0,  z_1, z_2, \cdots , z_{N+1}=F$, whose the lengths of steps are equal to $\Delta z$. 

For the time and the space second derivative, the forward  and the central difference schemes are employed, respectively.  Moreover, at any node, we employ the forward (backward) scheme for the first space derivative when the function $\alpha$ is nonnegative (negative).  
Therefore, denoting  $V(i,j)=V(t_i,z_j)$,  $0\leq i\leq M, 0\leq j\leq N+1$, the discretization of Equation (\ref{maineq}) at any node $(t_i,z_j), 0\leq i\leq M-1, 1\leq j\leq N$,  is given by
\begin{align}
\frac{V(i+1,j)-V(i,j)}{\Delta t}\nonumber\\
\quad \qquad \qquad +\inf_{\pi^1, \pi^2}\{& a(i,j) V(i,j+1)+b(i,j)V(i,j-1)\nonumber\\
&- \left(a(i,j)+b(i,j)\right)V(i,j)+\kappa \eta_{t_i} (C_1-\pi^2)^2\}=0, \quad \label{discrete}
\end{align}
in which 
$$a(i,j)=\frac{\alpha(t_i,z_j)}{\Delta z}+\frac{\beta(t_i,z_j)}{(\Delta z)^2}, \quad b(i,j)=\frac{\beta(t_i,z_j)}{(\Delta z)^2},\qquad \quad  when\; \; \alpha(t_i,z_j)\geq 0, $$ 
and  
$$a(i,j)=\frac{\beta(t_i,z_j)}{(\Delta z)^2}, \quad b(i,j)=\frac{\beta(t_i,z_j)}{(\Delta z)^2}-\frac{\alpha(t_i,z_j)}{\Delta z}, \qquad \quad when\; \; \alpha(t_i,z_j)<0. $$ 
The above representation clearly implies that the coefficients of $V(i,j+1)$ and $V(i,j-1)$ are nonnegative, which means that our scheme has the positive coefficient property; see  \cite[Condition 4.1]{kn:FL} for the more precise definition of this property. 
\begin{rem}
It should be noted that for the parameters that are considered in the next section and corresponding to the obtained optimal strategies, the function $\alpha$ is nonnegative at any point. 
\end{rem}

Moreover, the Lipschitz continuity of the functions $\alpha$, $\beta$ and smoothness property of the loss function satisfy the assumptions of \cite[V, Theorem 8.1]{kn:FS} which implies the comparison property for Equation  (\ref{maineq}). 

It is well known that  having the comparison property,  a numerical scheme with the positive coefficient property is convergent, if it is stable in $l_{\infty}$ norm, monotone  and consistent;  see \cite[Theorem 2.1]{kn:BS}. We show these properties in the following propositions. 

\begin{prop}\label{bound}
The discretization (\ref{discrete}) satisfies $l_{\infty}$- stability property, 
\begin{align}
\norm{V(i,\cdot)}_{\infty} \leq \left(\frac{F-S}{a_{a_1}}\right)^2+ \kappa (C_1-C_2)^2, \qquad  0\leq i\leq M.\label{stability}
\end{align}

\begin{proof}
By employing the optimal strategies $\pi^{1,*}, \pi^{2,*}$ in the coefficients of (\ref{discrete}), we have  for $0\leq i\leq M-1$ and $1\leq j\leq N$ 
\begin{align}
V(i,j)=&V(i+1,j)+\Delta t  a(i,j) V(i,j+1)+\Delta t b(i,j)V(i,j-1) \nonumber\\
&-\Delta t (a(i,j)+b(i,j)) V(i,j)\nonumber\\
&+\Delta t\kappa \eta_{t_i} (C_1-\pi^{2,*})^2.\label{eq}
\end{align}
So, we have               
\begin{align}
\abs{V(i,j)}\left(1+\Delta t(a(i,j)+b(i,j))\right)\leq&\norm{V(i,\cdot)}_{\infty}\Delta t \left(a(i,j)+b(i,j)\right)\nonumber\\
&+V(i+1,j)+\Delta t\kappa \eta_{t_i} (C_1-C_2)^2.\label{ineq}
\end{align}
If $V(i,j_1)=\norm{V(i,\cdot)}_{\infty}=max_{1\leq j\leq N}V(i,j)$, then considering the above inequality for the node $(i, j_1)$, we get 
\begin{align}
\norm{V(i,\cdot)}_{\infty}\left(1+\Delta t(a(i,j_1)+b(i,j_1))\right)\leq &\norm{V(i,\cdot)}_{\infty}\Delta t \left(a(i,j_1)+b(i,j_1)\right)  \nonumber\\
&+\norm{V(i+1,\cdot)}_{\infty}+\Delta t\kappa \eta_{t_i} (C_1-C_2)^2,\nonumber 
\end{align}
which means 
$$\norm{V(i,\cdot)}_{\infty}\leq  \norm{V(i+1,\cdot)}_{\infty}+\Delta t\kappa \eta_{t_i} (C_1-C_2)^2.$$
Now, regarding the terminal condition of $V$, the bound (\ref{stability}) is achieved. 

Furthermore, the bound is easily achieved on the upper and lower borders of the boundary. 
\end{proof}
\end{prop}

\begin{rem}\label{rem1}
The infimum in Equation (\ref{discrete}) indicates that by employing any admissible strategy in the coefficients, the equality in (\ref{eq}) turns into an inequality that still yields the inequality (\ref{ineq}). So, the bound (\ref{stability}) holds uniformly over the set of admissible strategies. 
\end{rem}

If we denote the left hand side of Equation (\ref{discrete}) by 
$$G^i_j\left( V(i,j), V(i,j+1), V(i,j-1), V(i+1,j)\right ),$$ 
the monotone property is stated in the following.  

\begin{prop}\label{monotone}
The discretization scheme (\ref{discrete}) is monotone,  that is for any $\eps_i \geq 0, \; i=1, 2, 3$ we have 
\begin{align}
G^i_j&\left(V(i,j), V(i,j+1)+\eps_1, V(i,j-1)+\eps_2, V(i+1,j)+\eps_3\right)\nonumber \\
&-G^i_j\left(V(i,j), V(i,j+1), V(i,j-1), V(i+1,j)\right )\geq 0. \nonumber
\end{align}
\begin{proof}
The coefficients of $V(i,j+1)$, $V(i,j-1)$ and $V(i+1,j)$ in (\ref{discrete}) are  all nonnegative. So, the above inequality is obtained directly.  
\end{proof}
\end{prop}

\begin{prop}\label{consistence}
The discretization scheme (\ref{discrete}) is consistent,  that is for any smooth test function $\phi:[0,T]\times [S,F]\rightarrow \Real$ with bounded derivatives of all orders with respect
to $t$ and $z$, and denoting $\phi^i_j=\phi(t_i, z_j)$,  we have 
\begin{align}
\lim_{\Delta t, \Delta z\rightarrow 0} \abs{ \left(\phi_t+\inf_{\pi^1, \pi^1}\{\mathcal{A}\phi+\kappa \eta_t (C_1-\pi^2)^2 \}\right)^i_j-G^i_j(\phi^i_j, \phi^i_{j+1}, \phi^i_{j-1}, \phi^{i+1}_j) }=0.\nonumber
\end{align}

\begin{proof}
Since the coefficients  $\alpha$ and $\beta$, and the derivatives of $\phi$ are all bounded on the domain $[0, T] \times [S, F]$, the following approximations are concluded by using Taylor series expansion 
$$\abs{ (\mathcal{A}\phi)^i_j -\left( \alpha(i,j) \frac{\phi^i_{j+1}-\phi^i_j }{\Delta z}+\beta(i,j) \frac{\phi^i_{j+1}-2\phi^i_j+\phi^i_{j-1}}{(\Delta z)^2}\right ) }=O(\Delta z),$$
$$\abs{ (\mathcal{A}\phi)^i_j -\left(- \alpha(i,j) \frac{\phi^i_{j-1}-\phi^i_j }{\Delta z}+\beta(i,j) \frac{\phi^i_{j+1}-2\phi^i_j+\phi^i_{j-1}}{(\Delta z)^2}\right ) }=O(\Delta z),$$
which correspond to the forward and backward schemes of the first order derivative, respectively. 

Moreover, clearly we have $\abs{(\phi_t)^i_j - \frac{\phi^{i+1}_j-\phi^i_j}{\Delta t} }=O(\Delta t) $, which concludes the proposition.  
\end{proof}
\end{prop}

\subsection{Policy iteration method}  
Since the coefficients in (\ref{discrete}) depend on the control variables, applying the implicit time stepping method yields nonlinear equation at each time step. To overcome this difficulty, at each time step, we apply the policy iteration method, in which the optimal strategy and the value function are approximated iteratively.
 
The value function is known at the terminal time $T$. So, we go backward in time and start  from the last column $t=T-\Delta t$ inside the domain $\mathcal{C}^{\prime}$. On each column $t=t_i$, we take as the starting point the value function on the next time step, the vector $V(t_i+\Delta t, \cdot)$, and apply the following algorithm.  \\ 

\textbf{Policy iteration algorithm:}

\textbf{I)} \emph{Policy improvement:} For the given value function on the nodes $(t_i,z_j), 1\leq j\leq N$,  solve the static optimization problem (\ref{discrete}) to find optimal investment and consumption strategies $\pi^{1}$ and $\pi^{2}$, respectively.  

\textbf{II)} \emph{Policy evaluation:} Employing the strategies $\pi^1, \pi^2$ obtained from Step (I) and considering the boundary conditions (\ref{boundary})-(i)-(ii)-($iii^{\prime}$), solve the linear system (\ref{discrete}) corresponding to all nodes $(t_i,z_j), 1\leq j\leq N$,  to find new value function, $V^{new}(t_i,\cdot)$, on the column $t=t_i$.  

\textbf{III)} \emph{Convergence criterion:}  If the following inequality does not hold, return to the Step (I)   
$$\max_{1\leq j\leq N} \abs{V^{new}(t_i,z_j)-V^{old}(t_i,z_j)}\leq \{ \max_{1\leq j\leq N} \abs{V^{new}(t_i,z_j)} \}\times 10^{-6}. $$

In the next theorem, we prove that the iteration policy gives a monotone sequence of approximate value functions.  Our proof is a modification of the proof given in \cite[Theorem 1]{kn:PFV}.  

We write the discrete representation (\ref{discrete}) in the matrix form. Let $V^i=V(i,\cdot)$ be the vector represents the value function on $t=t_i$ and $D^i$ the $(N+2)\times (N+2)$ tridiagonal matrix in which the first and last rows are zero and 
\begin{align}
[D^iV^i]_j=&a(i,j) V^i(j+1)+b(i,j)V^i(j-1)\nonumber\\ 
&-\left(a(i,j)+b(i,j)\right)V^i(j), \qquad \qquad \qquad 1\leq j\leq N. \nonumber
\end{align}
Then we can rewrite (\ref{discrete}) for all $0\leq i\leq M-1$ as 
\begin{align}
[I-\Delta t D^i]V^i=V^{i+1}+\Delta t E^i+G^i-G^{i+1}, \label{vec}
\end{align}
in which $E^i=\kappa \eta_{t_i} (C_1-\pi^2)^2$ and $G^i$ is a vector with just one nonzero element stands for the boundary value of $V$ at $(t_i,S)$.  Notice that $(I-\Delta t D^i)$ is a tridiagonal matrix in which the diagonal entries are positive, the off-diagonal entries are negative and the row sums are all equal to 1. So, it is an $M$-matrix and therefore we have $(I-\Delta t D^i)^{-1} \geq 0$. 

\begin{thm}
For any fixed column $t=t_i$, the policy iteration algorithm gives a monotone sequence of vectors which converges to $V^i$.

\begin{proof}
Let $W^0=V^{i+1}$  and $W^k, k\geq 1$ be the approximation of $V^i$ obtained in Step (II) in the kth iteration. Then, from (\ref{vec}) we have 
\begin{align}
[I- \Delta t D^{i,k}(\pi^{1,k}, \pi^{2,k})] W^{k}=V^{i+1}+\Delta t E^{i,k}(\pi^{1,k}, \pi^{2,k})+G^i-G^{i+1}, \label{iteration}
\end{align}
where, for any $k\geq 1$,
\begin{align}
(\pi^{1,k}, \pi^{1,k})=arg min_{(\pi^{1,k}, \pi^{2,k})\in \Pi_{ad}} \left \{ D^{i,k}(\pi^{1,k}, \pi^{2,k}) W^{k-1}+E^{i,k}(\pi^{1,k}, \pi^{2,k})\right \}. \label{strat}
\end{align}
Hence, the coefficients $D^{i,k}, E^{i,k}$ are modified in each iteration. 

In view of (\ref{iteration}) and doing some manipulations, we get 
\begin{align}
[I- \Delta t D^{i,k+1}] (W^{k+1}-W^k)=\Delta t \left [(D^{i,k+1}W^k+E^{i,k+1})-(D^{i,k}W^k+E^{i,k}) \right ].\label{diff}
\end{align}
The construction (\ref{strat}) implies that, given $W^k$, the coefficients $D^{i,k+1}$ and $E^{i,k+1}$ yield the minimum value for all elements of the vector  $(D^{i,k+1}W^k+E^{i,k+1})$. Therefor we have for all $0\leq j\leq N+1$
$$\left [(D^{i,k+1}W^k+E^{i,k+1})-(D^{i,k}W^k+E^{i,k}) \right ]_j \leq 0. $$
Now, since $(I-\Delta t D^{i,k+1})^{-1} \geq 0$, we conclude from (\ref{diff}) that the sequence of vectors $W^k, k\geq 0$ is decreasing.  Furthermore, due to Remark \ref{rem1}, the sequence $(W^k)_{k\geq 0}$ is uniformly bounded. So, it is convergent. 

The construction process of $W^k$ implies that it converges to $V^i$, the unique solution of (\ref{discrete}), which is an approximation of the value function at  $t=t_i$. 
\end{proof}
\end{thm}

\section{\textbf{Simulation Results }}
For comparison purposes, we assume the same market parameters as in \cite{kn:Ger1} and \cite{kn:G}. So, we assume the interest rate to be $r=0.03$ and the expected return and the volatility of the risky asset $\mu=0.08$ and $\sigma=0.15$, respectively, which implies a Sharpe ratio equal to $\beta=0.33$. Furthermore, we consider a retiree with age $a_0=60$ and initial wealth $x_0=100$. In addition, the length of the decumulation phase is assumed to be equal to $T=15$ years, which means that $a_1=75$. The maximum consumption rate is set to be $C_1=6.5155$, which equals the payments of a lifetime annuity purchasable at the retirement time, in view of the mortality rate given in this section. 

We consider four scenarios for the minimum admissible consumption rate, $C_2=C_1$, $C_2=\frac{3}{4}C_1$, $C_2=\frac{2}{3}C_1$ and $C_2=\frac{1}{2}C_1$. Moreover, the target level $F=1.75C_1 a_{75}$ and the safety level $S=0.5 C_1a_{75}$ are considered for the wealth process, which in the literature correspond  to the medium level of risk aversion; see  \cite{kn:Ger1} and \cite{kn:G}. This means that in our setting, the final annuity that the retiree will receive is at most 1.75 and at least 0.5 times $C_1$. It is clear that the higher the risk aversion, the higher the target $F$, and the lower the safety level $S$.  
 
 We should note that our main concern in this work is comparing the different scenarios of the admissible ranges of consumption. So, here the levels $F$ and $S$ are fixed. However, our simulation results corresponding to the other levels of risk aversion, which have not been reported here, demonstrate results similar to the ones obtained for the medium level of risk aversion. 
   
\begin{figure}
  \begin{multicols}{2}
    \hspace{-4cm}
    \begin{minipage}{0.9\textwidth}
        \centering
        \includegraphics[width=0.85\textwidth]{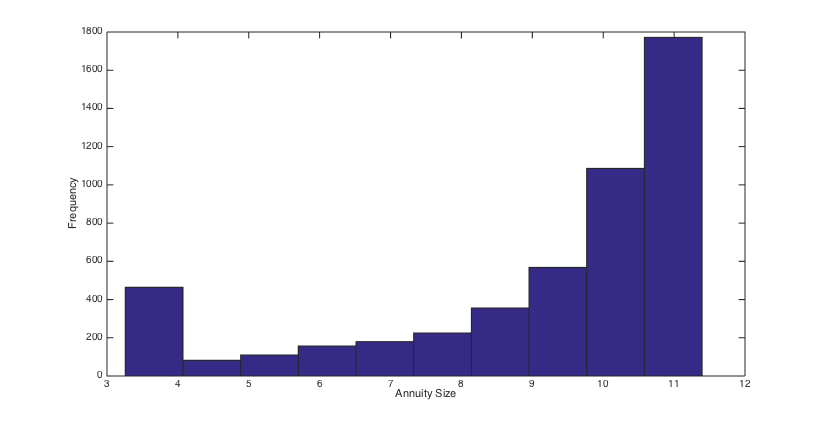} 
        \caption{ \footnotesize{Final Annuity ($C_2=\frac{1}{2}C_1$)}}
        \label{fig1}
    \end{minipage}

\vspace{1cm}
 \hspace{-4cm} 
\begin{minipage}{0.9\textwidth}
        \centering
        \includegraphics[width=0.85\textwidth]{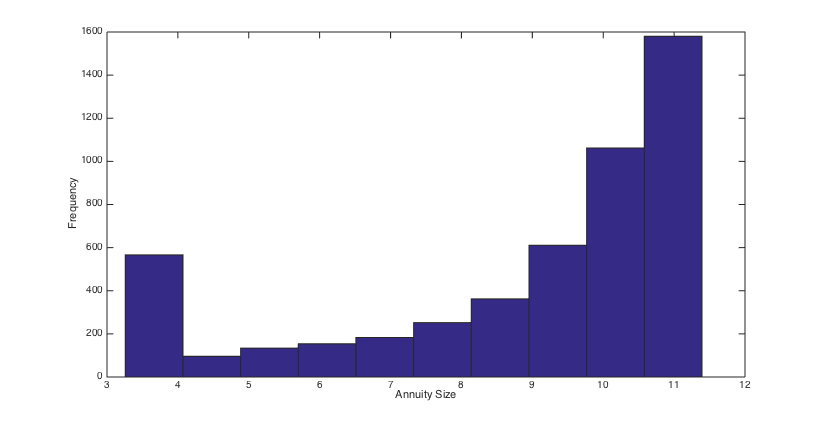} 
        \caption{ \footnotesize{Final Annuity ($C_2=\frac{3}{4}C_1$)}}
        \label{fig2}
    \end{minipage}


\vspace{1cm}
  \hspace{-4cm}
    \begin{minipage}{0.9\textwidth}
        \centering
        \includegraphics[width=0.85\textwidth]{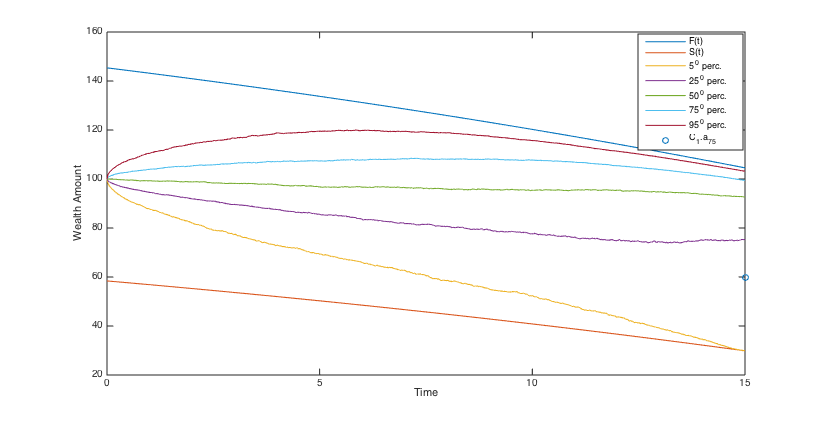} 
        \caption{ \footnotesize{Optimal Wealth ($C_2=\frac{1}{2}C_1$)}}
        \label{fig3}
    \end{minipage}

 \hspace{-1.5cm} 
    \begin{minipage}{0.9\textwidth}
            \centering
        \includegraphics[width=0.85\textwidth]{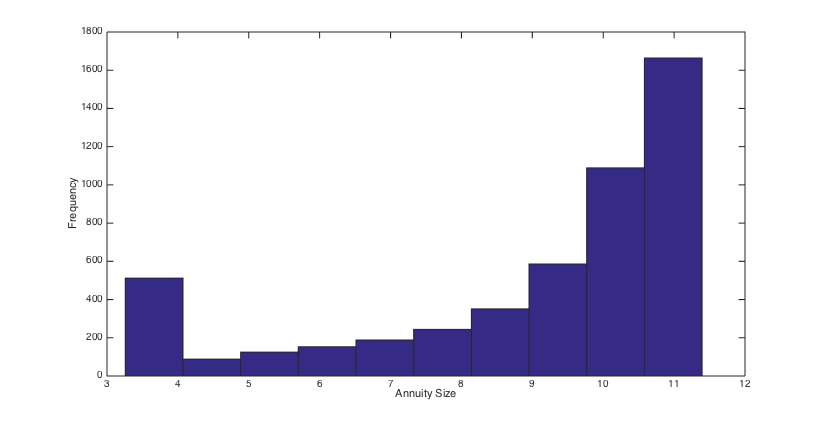} 
        \caption{ \footnotesize{Final Annuity ($C_2=\frac{2}{3}C_1$)}}
        \label{fig4}
    \end{minipage}    

\vspace{1cm}
 \hspace{-1.5cm} 
    \begin{minipage}{0.9\textwidth}
        \centering
        \includegraphics[width=0.85\textwidth]{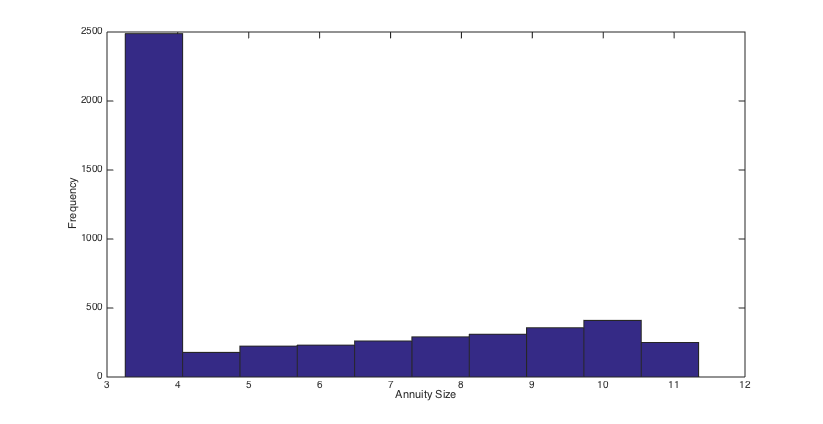} 
        \caption{ \footnotesize{Final Annuity ($C_2=C_1$)}}
        \label{fig5}
    \end{minipage}

\vspace{1cm}
 \hspace{-1.5cm} 
    \begin{minipage}{0.9\textwidth}
        \centering
        \includegraphics[width=0.85\textwidth]{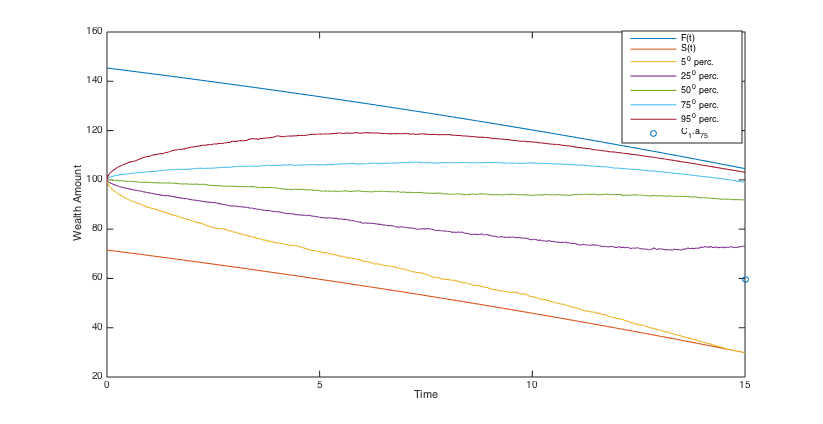} 
        \caption{ \footnotesize{Optimal Wealth ($C_2=\frac{2}{3}C_1$)}}
        \label{fig6}
    \end{minipage}

 \end{multicols}
\end{figure}

\begin{figure}
 \begin{multicols}{2}

\vspace{-4.5cm}
 \hspace{-4cm}
    \begin{minipage}{0.9\textwidth}
        \centering
        \includegraphics[width=0.85\textwidth]{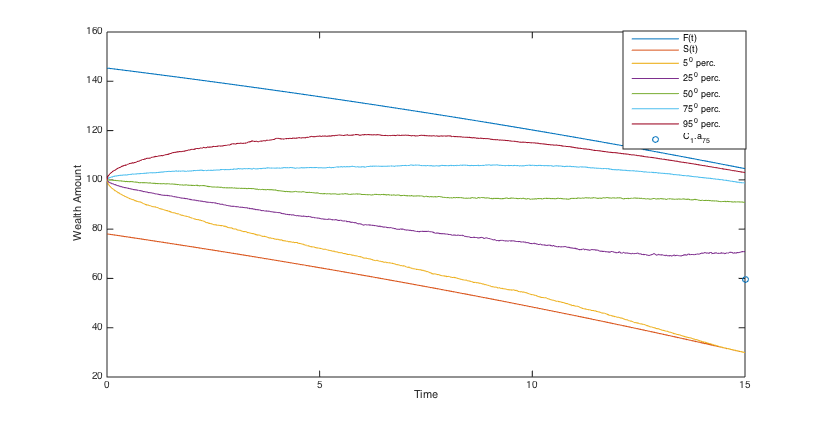} 
        \caption{ \footnotesize{Optimal Wealth ($C_2=\frac{3}{4}C_1$)}}
        \label{fig7}
    \end{minipage}

\vspace{1cm}
 \hspace{-4cm}
\begin{minipage}{0.9\textwidth}
        \centering
        \includegraphics[width=0.85\textwidth]{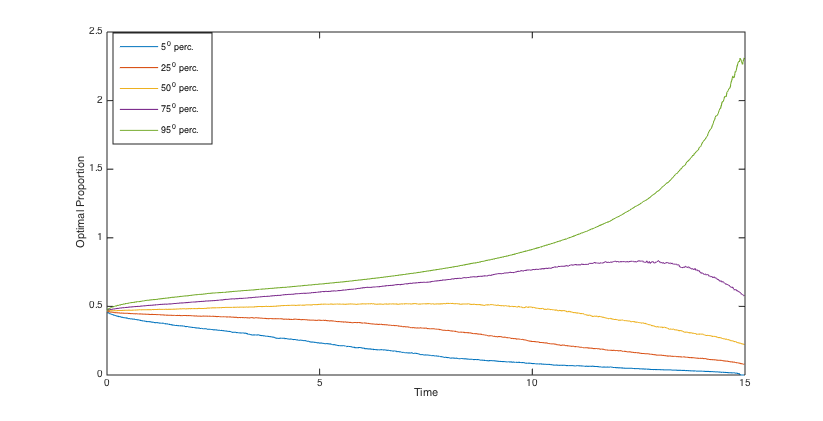} 
        \caption{ \footnotesize{Risky Investment ($C_2=\frac{1}{2}C_1$)}}
        \label{fig8}
    \end{minipage}

\vspace{1cm}
 \hspace{-4cm}
 \begin{minipage}{0.9\textwidth}
        \centering
        \includegraphics[width=0.85\textwidth]{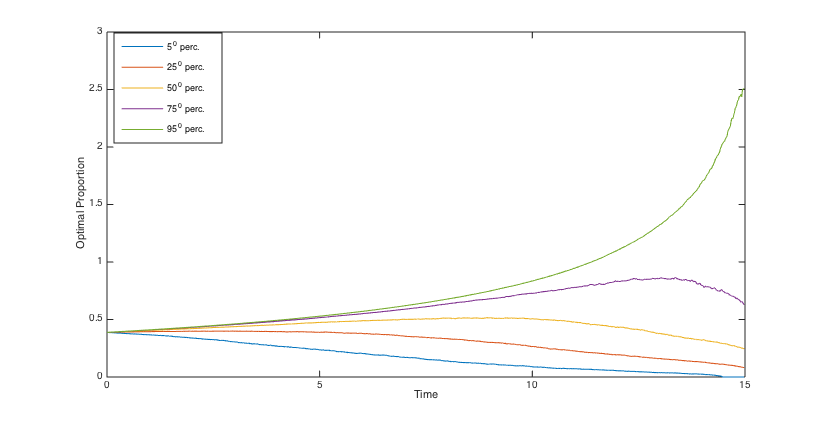} 
        \caption{ \footnotesize{Risky Investment ($C_2=\frac{3}{4}C_1$)}}
        \label{fig9}
    \end{minipage}
   

 \hspace{-1.5cm}
    \begin{minipage}{0.9\textwidth}
        \centering
        \includegraphics[width=0.85\textwidth]{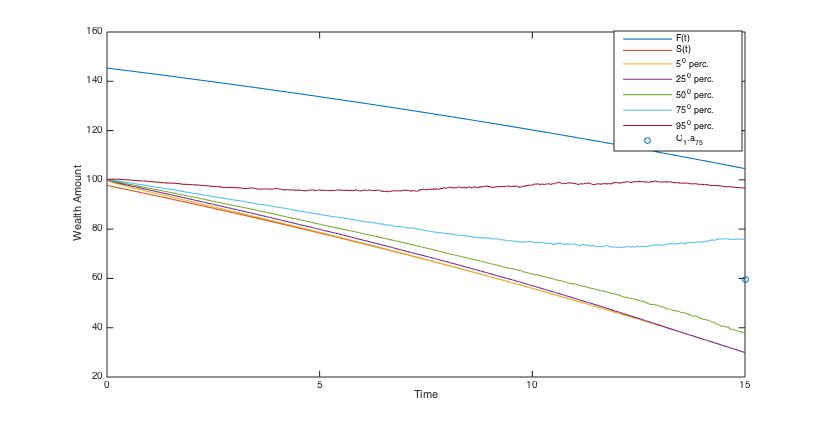} 
        \caption{ \footnotesize{Optimal Wealth ($C_2=C_1$)}}
        \label{fig10}
    \end{minipage}

\vspace{1cm}
 \hspace{-1.5cm}
    \begin{minipage}{0.9\textwidth}
        \centering
        \includegraphics[width=0.85\textwidth]{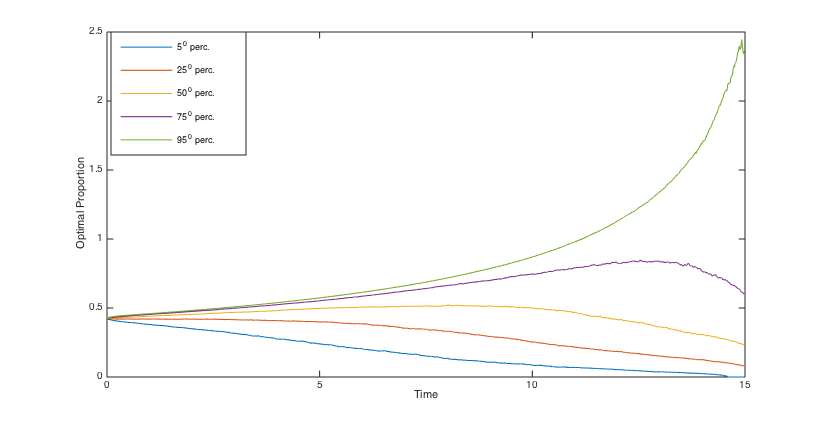} 
        \caption{ \footnotesize{Risky Investment ($C_2=\frac{2}{3}C_1$)}}
        \label{fig11}
    \end{minipage}

  \vspace{1cm}
   \hspace{-1.5cm}
     \begin{minipage}{0.9\textwidth}
        \centering
        \includegraphics[width=0.85\textwidth]{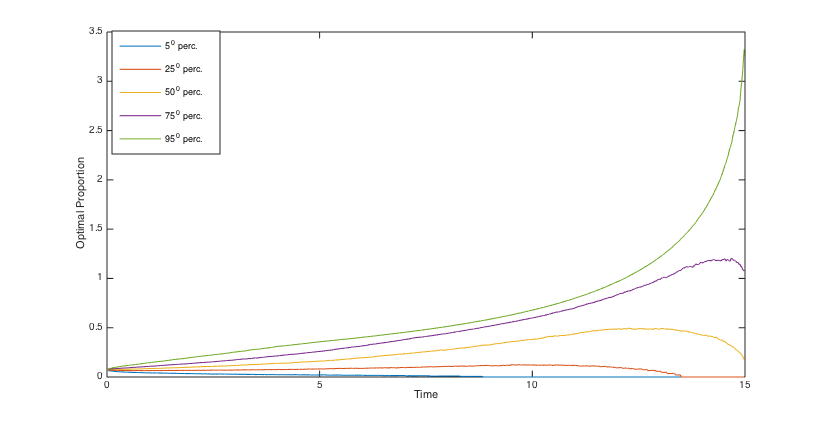} 
        \caption{ \footnotesize{Risky Investment ($C_2=C_1$)}}
        \label{fig12}
    \end{minipage}

 \end{multicols}
\end{figure}

To simulate the optimal wealth process, the same stream of 5000 pseudo random numbers is applied to different scenarios.  
Using the simulated optimal wealth processes when $\kappa=0.5$, we present the histograms of the final annuities in Figs. \ref{fig1}, \ref{fig2}, \ref{fig4} and \ref{fig5}  and reveal some percentiles  of the optimal wealth amounts in Figs. \ref{fig3}, \ref{fig6}, \ref{fig7} and \ref{fig10}, and the optimal investment strategies in Figs. \ref{fig8}, \ref{fig9}, \ref{fig11} and \ref{fig12}.

The results show that when the admissible range of consumption is more restricted the investment is a little more risky and the final annuity is higher, on average. Moreover, the main indication of the graphs and histograms is the big difference between the final annuity and the optimal wealth process that correspond to the fixed consumption scenario and those that correspond to three other scenarios. Actually, we conclude from the results  that by assuming a variable consumption rate, although quite restricted to a short range, we get considerably more valuable final annuities and  greater optimal wealth amounts. 

\begin{figure}
 \begin{multicols}{2}
 
\vspace{-4.5cm}
 \hspace{-4cm}
    \begin{minipage}{0.9\textwidth}
        \centering
        \includegraphics[width=0.85\textwidth]{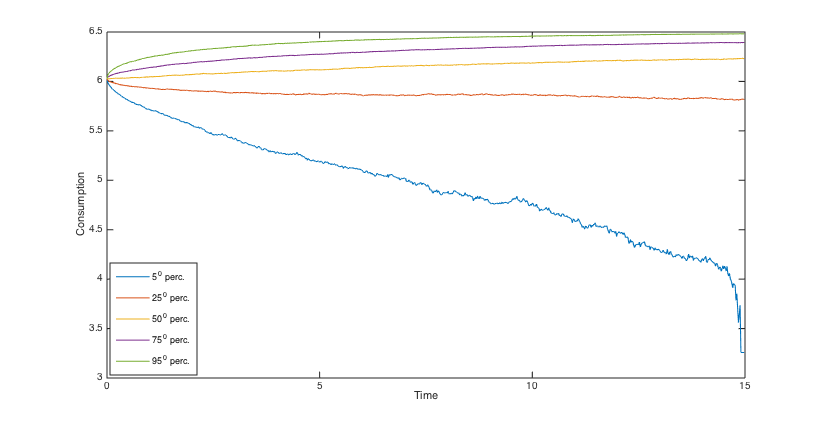} 
        \caption{ \footnotesize{Optimal Consumption ($C_2=\frac{1}{2}C_1, \; K=0.5$)}}
        \label{fig13}
    \end{minipage}
    

\vspace{1cm}
  \hspace{-4cm}
\begin{minipage}{0.9\textwidth}
        \centering
        \includegraphics[width=0.85\textwidth]{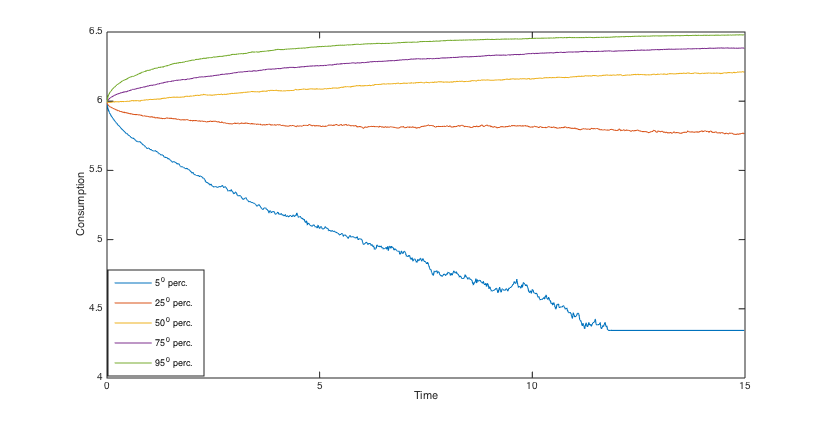} 
        \caption{ \footnotesize{Optimal Consumption ($C_2=\frac{2}{3}C_1, \; K=0.5$)}}
        \label{fig14}
    \end{minipage}
    
 \vspace{1cm}
  \hspace{-4cm}
 \begin{minipage}{0.9\textwidth}
     \centering
        \includegraphics[width=0.85\textwidth]{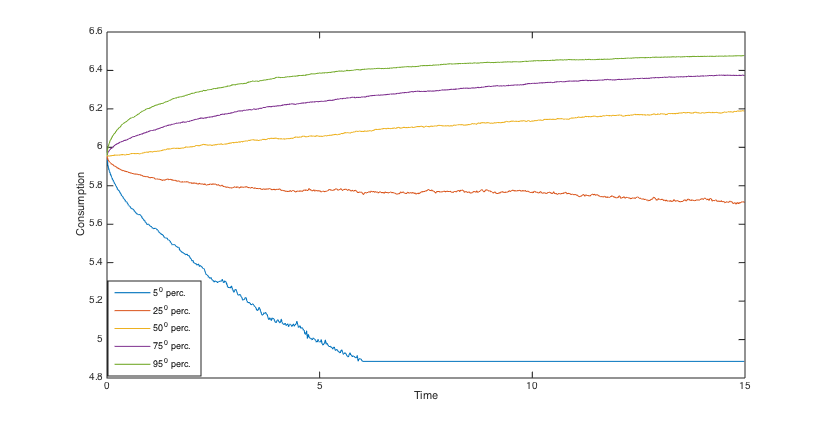} 
        \caption{ \footnotesize{Optimal Consumption ($C_2=\frac{3}{4}C_1, \; K=0.5$)}}
        \label{fig15}
    \end{minipage}
    

 \hspace{-1.5cm}
  \begin{minipage}{0.9\textwidth}
        \centering
        \includegraphics[width=0.85\textwidth]{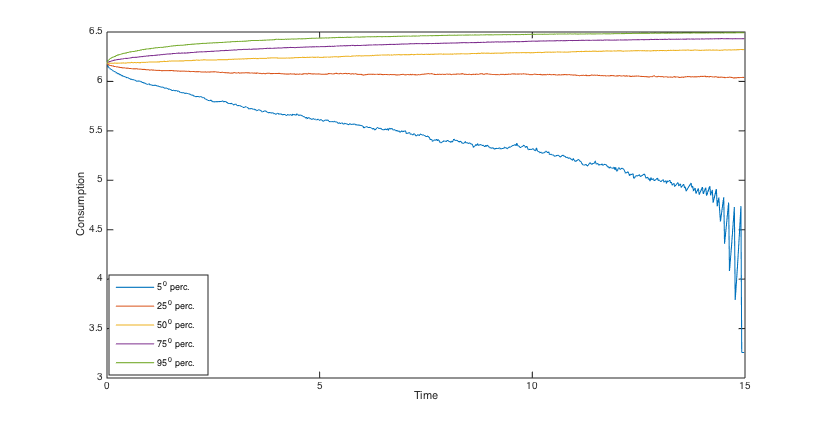} 
        \caption{ \footnotesize{Optimal Consumption ($C_2=\frac{1}{2}C_1,\; K=1$)}}
        \label{fig16}
    \end{minipage}

\vspace{1cm}
 \hspace{-1.5cm}
  \begin{minipage}{0.9\textwidth}
        \centering
        \includegraphics[width=0.85\textwidth]{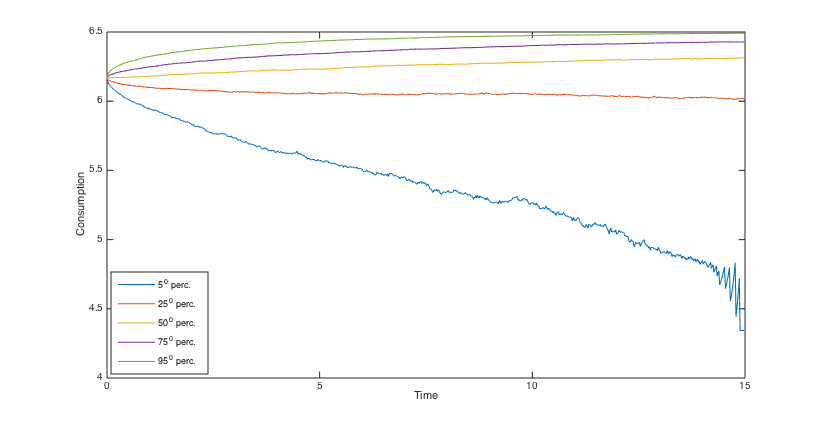} 
        \caption{ \footnotesize{Optimal Consumption ($C_2=\frac{2}{3}C_1,\; K=1$)}}
        \label{fig17}
    \end{minipage}
 
\vspace{1cm}
 \hspace{-1.5cm} 
  \begin{minipage}{0.9\textwidth}
        \centering
        \includegraphics[width=0.85\textwidth]{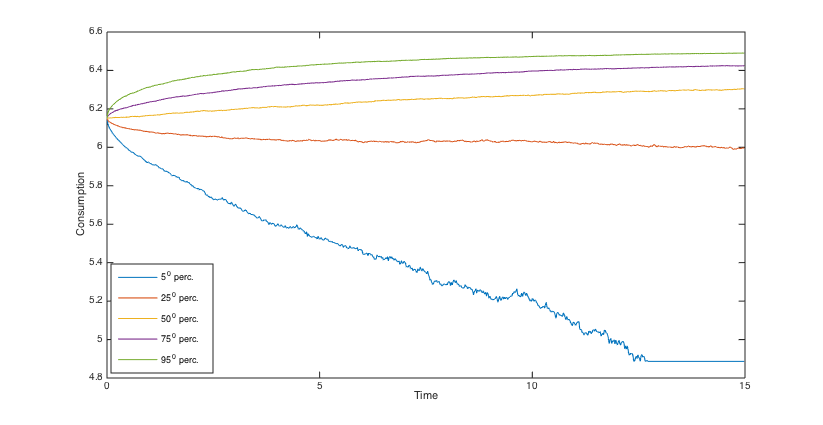} 
        \caption{ \footnotesize{Optimal Consumption ($C_2=\frac{3}{4}C_1,\; K=1$)}}   
        \label{fig18}       
    \end{minipage}   

 \end{multicols}
\end{figure}

The percentiles of the optimal consumption rate for $\kappa=0.5$ and $\kappa=1$, are exhibited in Figs. \ref{fig13}-\ref{fig15},  and in Figs. \ref{fig16}-\ref{fig18}, respectively. We see that compared to $C_1=6.5155$, the consumption rates  are fifty percent greater than $6$, and therefore are not much smaller than $C_1$. Moreover, as it is expected, the optimal consumption is higher when $\kappa$ is greater. 

For a comprehensive comparison of the outcomes of different scenarios, we take into account  the market present value of the cash flows before and after the annuitization. The cash flow before the annuitization consists of withdrawals $\pi^2_t$ from the fund, for  $0\leq t\leq T$. Moreover, in the case of death before the annuitization, we also take into account the accumulated wealth at the time of death. The cash flow after the annuitization consists of the constant payments of the lifetime annuity that is purchased by the accumulated wealth at the terminal time $T$. 
Therefore, considering $X_t$ as the fund value at time $t$, the present value of the cash flows is written as 
\begin{align}
P.V.=\int^{\tau_d\wedge15}_0e^{-\rho t }\pi^2_t dt+e^{-\rho\tau_d}X_{\tau_d} 1_{\{\tau_d \leq 15\}} +\int^{\tau_d}_{ \tau_d\wedge 15}  \frac{X_{15}}{a_{75}}e^{-\rho t }dt,\nonumber
\end{align}
in which $\tau_d=\tilde{\tau}_d-60$, where $\tilde{\tau}_d$ denotes the time of death.  

For the mortality rate, the Gompertz-Makeham distribution is considered, which for an individual of age $t$ is given as  
$$\nu_t=A+BC^t.$$
The parameters $A, B$ and $C$ are those considered by the Belgian regulator for the pricing of the lifetime annuities purchased by males, as in \cite{kn:HD}. So, we assume $A=0.00055845$, $B=0.000025670$ and $C=1.1011$.

Now, assuming  that the time of death is independent of the filtration of financial returns, we get
\begin{align}
P.V.=\int^{15}_0\eta_t \left(\pi^2_t +\nu_{60+t}X_t\right) dt+\int^{T_m-60}_{15} \eta_t \frac{X_{15}}{a_{75}} dt, \nonumber
\end{align}
in which $T_m=100$ is considered  as the  maximum lifespan. 

Table \ref{tab2} reveals some statistics of  the final annuity and the present value of the cash flows for different admissible ranges of consumption and different weights of the running cost, $\kappa=0.25, 0.5, 0.75,$ and 1. Furthermore, it reports the mean of the consumptions before the annuitization and the probability that the final annuity is greater than $C_1$, the annuity that is  purchasable at the retirement time. 

As it is expected, when the minimum admissible consumption rate $C_2$ decreases or equivalently the admissible range of consumption becomes wider, we get more valuable final annuity and higher cash flows. Furthermore, in this case the probability of achieving a final annuity greater than $C_1$ increases.  

We see in Table \ref{tab2} again the remarkable difference between the outcomes of the fixed consumption rate scenario $C_2=C_1$ and the outcomes that correspond to three other scenarios. Actually,  when the consumption rate is fixed, considerably less valuable final annuity is achieved. Furthermore, the present value of the cash flows is remarkably less than the  analogous outcomes of the variable consumption rate scenarios. However, since in this  case the cash flow before the annuitization is constant, the total cash flows become less deviated.  


\begin{table} 
\vspace{-1.2cm}
\hspace{-2cm}

\caption{\footnotesize{Distribution of final annuities, consumptions and cash flows}} 
\vspace{0.5cm}
\begin{tabular}{c  c  c  c  c  c}
\hline 
&\footnotesize{ $C_2 =\frac{1}{2}C_1$}
&\footnotesize{$C_2=\frac{2}{3}C_1$} 
&\footnotesize{ $C_2=\frac{3}{4}C_1$}
&\footnotesize{ $C_2=C_1$}\\
\hline
&$\kappa=0.25$  &$\kappa=0.25$  &$\kappa=0.25$   \\
&$\kappa= 0.5$  &$\kappa=0.5$  &$\kappa=0.5$    \\
&$\kappa=0.75$  &$\kappa=0.75$ &$\kappa=0.75$    \\
&$\kappa= 1$  &$\kappa=1$  &$\kappa=1$     \\
\hline
\hline
Mean of FA\footnotemark                      &9.59                   &9.43                          &9.28                        &5.69 \\
                                                             &9.08                     &8.96                         & 8.85                                      \\ 
                                                             &8.77                      &8.67                        &8.57                           \\
                                                              &8.54                     &8.45                       & 8.37                                \\
\hline
Standard deviation of FA                                                &2.14                    &2.24                             &2.33                      &2.77   \\
                                                            &2.43                       & 2.50                           &2.55                                       \\
                                                            &2.58                       &2.63                            &2.67                                      \\
                                                            &2.68                       & 2.71                           &2.74                                   \\
\hline
Mean of PV\footnotemark                               &105.21                                 &104.15                              &103.25          &96.17   \\
                                                                   &105.34                                &104.59                              & 103.92                           \\
                                                                   &105.06                                 &104.46                              &103.91                           \\
                                                                   &104.71                               & 104.22                                 & 103.76                            \\

\hline
Standard deviation of PV                                                           &13.95                              &13.77                              &13.49            &12.17   \\
                                                                    &14.11                                &14.14                             & 14.08                             \\
                                                                    &14.12                                &14.19                              &14.21                           \\
                                                                     &14.16                               &14.18                                & 14.21                         \\
\hline
\footnotesize{Prob(FA $>  C_1$) (\%)}                                    &89.62                          &88.10                      &86.22       &37.44  \\
                                                                                                &83.72                          &82.42                      & 80.98                       \\
                                                                                                &80.10                           &79.08                     &77.92                   \\
                                                                                                &77.42                           & 76.54                    & 75.76                  \\
\hline
\footnotesize{Prob($X_T=S$)} (\%)                         &1.62                &1.94                         &2.16                 &6.94     \\
                                                                                                   &1.66                &2.60                         & 2.86                              \\
                                                                                                   &1.44                 &1.94                         &3.24                            \\
                                                                                                   &0.94               &2.18                          &3.26                                 \\ 
\hline 
Mean of consumptions                                                                   &5.7752              & 5.7466                    &5.7341                 &6.5155             \\
                                                                                                   &5.9869              &5.9698                     &5.9595                                              \\
                                                                                                   &6.0873               &6.0756                    &6.0668                                            \\
                                                                                                   &6.1470               &6.1398                     &6.1330                                    \\
\hline                                                                                                                     
$^1$FA=Final Annuity\\
 $^2$ PV=Present Value
\end{tabular}  
 \label{tab2}
\end{table}

For a fixed admissible range of consumption, comparing the outcomes that correspond to different weights of the running cost, we find quite reasonably, finer final annuities (higher mean with lower standard deviation) for smaller weights. Actually, when $\kappa$ is smaller, more weight is devoted to the second term of the loss function, a term which is based on the final annuity. 

The surprising result for a fixed admissible range of consumption is that the maximum present value of cash flows is attained when $\kappa=0.5$. This can be interpreted by considering simultaneously  the two terms of the loss function. In fact, for a smaller $\kappa$, in which more weight of the loss function is devoted to the final annuity,  higher cash flow after the annuitization is expected. On the other hand, for a greater $\kappa$, in which  more weight is devoted to the consumption,  higher cash flow before the annuitization  is expected. 

The last rows of the table report the mean of the consumptions. As it is expected, we have on average more consumption for the greater $\kappa$. The surprising point is that when the admissible range of consumption becomes more limited, or $C_2$  increases, the mean of the consumptions slightly decreases. This phenomenon can be interpreted from the results obtained for the probability of hitting the lower border, $Prob\{X_T=S\}$, since this probability is higher when the range is more limited. Furthermore, whenever the wealth process hits the lower border, afterwards the consumption rate must be at the minimum rate.

\section{\textbf{Conclusions }}
The optimal investment-consumption problem post-retirement is investigated. Considering a minimum guarantee for the final annuity, a variable consumption rate together with a running cost term of loss function, which is based on the consumption, yield a nonlinear HJB equation on an irregular bounded domain. Using the dual equation, the existence and uniqueness of a classical solution are proved.  To obtain a numerical approximation of the solution of the HJB equation, we apply the backward in time implicit scheme of the finite difference method. Our scheme has the positive coefficient property which guarantees  convergence to the viscosity solution. To tackle the nonlinearity that appears in the numerical scheme, the policy iteration method is applied. 

Four different scenarios for the admissible range of consumption are considered. Our simulation results show that by assuming a variable consumption rate, although quite restricted,  a much finer final annuity is achieved. Furthermore, the outcomes corresponding to different weights of the running cost term are compared. 

Moreover, by evaluating the present value of the total cash flows, before and after the annuitization, we observe higher values when the admissible range of consumption is wider. 

Our goal for future research is to consider a market model with jump-diffusion dynamics, which seems to be a challenging problem. 


\section{\textbf{Appendix}}

\textbf{Proof of Proposition \ref{con}}

We follow the steps of proof of the same statement in \cite[Proposition 4.8]{kn:DG}. Since in our problem there are two control variables and the running cost term in the loss function is based on the consumption, our proof is a little different from the corresponding proof in \cite{kn:DG}. The steps 2, 3 and 5 in the existing proof can be applied to our problem without essential modifications. So, here we just prove the steps 1 and 4. \\

\emph{Step 1.} For any fixed $S\leq z\leq F$, we prove that the function 
\begin{align}
t\rightarrow V(t,z)+(C_1-C_3)^2 t, \label{tfunc}
\end{align}
 in which $C_3=C_1+(C_2-C_1)\frac{F-z}{F-S}$, is nondecreasing. Consider the function 
 $$R:[0, T]\rightarrow [S, T], \qquad  t\rightarrow R(t)=\frac{C_3}{r}+(z-\frac{C_3}{r})e^{-r(T-t)}.$$   
At first, we show that 
\begin{align}
\mathcal{L}_1(t,R(t))=z,  \qquad \qquad 0\leq t\leq T. \label{C_3}
\end{align}
A few manipulations yield 
 \begin{align}
 F(t)-R(t)&=(F-z)\left [\frac{(C_1-C_3)(1-e^{-r(T-t)})}{r(F-z)}+e^{-r(T-t)}\right ]\nonumber\\
 &=(F-z)\left [\frac{(C_1-C_2)(1-e^{-r(T-t)})}{r(F-S)}+e^{-r(T-t)}\right ],\nonumber
 \end{align}
and
 \begin{align}
 F(t)-S(t)&=(F-S)\left [\frac{(C_1-C_2)(1-e^{-r(T-t)})}{r(F-S)}+e^{-r(T-t)} \right ].\nonumber
 \end{align}
 So, we have $\frac{F(t)-R(t)}{F(t)-S(t)}=\frac{F-z}{F-S}$, and consequently 
 \begin{align}
 \mathcal{L}_1(t,R(t))&=R(t)e^{r(T-t)}+\left [C_1+(C_2-C_1)\frac{F(t)-R(t)}{F(t)-S(t)}\right ]\frac{1-e^{r(T-t)}}{r}\nonumber\\
 &= \frac{C_3}{r}e^{r(T-t)}+ (z-\frac{C_3}{r})+C_3\frac{1-e^{r(T-t)}}{r}=z. \nonumber 
 \end{align}
 
Moreover, it can be checked easily that by applying the strategies $\pi^1=0$ and $\pi^2=C_3$, the wealth process will have the same dynamics as the function $R$. Therefore, if at any time $0\leq t\leq T$ the wealth amount equals $R(t)$, then it will remain on the curve $\{R(s), t\leq s\leq T\}$.  
 
 The above observations imply that the wealth process can move from the point $(t, z)$ to the point $(t^{\prime}, z)$ by the strategies which yield the cost amount $(C_1-C_3)^2(t-t^{\prime})$.  This yields  that the function (\ref{tfunc}) is nondecreasing.\\

\emph{Step 2.} The function $t\rightarrow V(t,z)$  is continuous for every $S\leq z\leq F$. \\

\emph{Step 3.}  The function $V$ is continuous on the set $[0, T] \times [S +\eps, F]$, for any $\eps > 0$. \\
 
 \emph{Step 4.}  In this step, we claim the right continuity of the value function $V$ in the space variable at $z=S$, for any fixed $0\leq t_0\leq T$. We prove the right continuity of $\tilde{V}(t_0, \cdot)$ at $x=S(t_0)$, which trivially implies the claim. 

Let at time $ t_0$ the wealth process be equal to $x_0=S(t_0)+\eps$, in which $\eps$ is a positive constant.  By applying the consumption rate $\pi^2\equiv C_2$ and the null investment strategy $\pi^1\equiv 0$, the wealth process will remain on the curve 
$$L(t)=\left (x_0-\frac{C_2}{r}\right )e^{r(t-t_0)}+\frac{C_2}{r},\qquad\quad t_0\leq t\leq T. $$
So, by applying the  strategy $\pi^2\equiv C_2$ and any admissible investment strategy $\pi^1$, the process $e^{-rt} \left(X(t; t_0, x_0, \pi^1(\cdot), C_2)-L(t)\right ), \; t_0\leq t\leq T,$ is a martingale under the risk-neutral measure. Therefore, we get the following equality for the expectation of the terminal wealth under the risk-neutral measure   
\begin{align}
\tilde{\mathbb{E}}X(T; t_0, x_0, \pi^1(\cdot), C_2)= L(T)=S+\eps e^{r(T-t_0)},\label{KK}
\end{align}
which converges to the safety level $S$ as $\eps \rightarrow 0$.

Then, by using the argument that is employed in \cite[Proposition 4.8, Step 4]{kn:DG}, we get the following convergence under the actual measure uniformly over all admissible investment strategies 
\begin{align}
 \mathbb{E} \left (F- X(T; t_0, x_0, \pi^1(\cdot), C_2)\right )^2\rightarrow (F-S)^2  \quad as \; \eps \rightarrow 0 . \label{first} 
 \end{align}
Moreover, it is clear that when $\pi^2\not\equiv C_2$, the terminal wealth amount is less than $X(T; t_0, x_0, \pi^1(\cdot), C_2)$. So, the above convergence holds uniformly over all admissible strategies $(\pi^1, \pi^2) \in \tilde{\Pi}_{ad}(t_0, x_0)$. 

Now, we are going to estimate the running cost loss function. We prove the following convergence for an admissible consumption strategy $\pi^{2,\eps}$    
\begin{align}
(T-t_0)(C_1-C_2)^2-\int^T_{t_0} (C_1-\pi^{2,\eps}_t)^2dt\rightarrow 0, \quad as \; \eps \rightarrow 0. \label{two}
\end{align}
Obviously, to prove the above convergence, it is enough to prove 
\begin{align}
\int^T_{t_0} (\pi^{2,\eps}_t-C_2)dt\rightarrow 0, \quad as \;\eps \rightarrow 0.\label{consumptionconver}
\end{align}
For an admissible strategy $(\pi^{1,\eps},\pi^{2,\eps})\in \tilde{\Pi}_{ad}(t_0, x_0)$, let 
$$Y(t)=\tilde{\mathbb{E}} X(t;t_0, x_0, \pi^{1,\eps}(\cdot), \pi^{2,\eps}(\cdot)),$$ 
be the expectation of the wealth process under the risk-neutral measure. Since $\pi^{2,\eps} \geq C_2$, we have $L(t)\geq Y(t)$, for all $t_0\leq t\leq T$. Regarding this inequality and the dynamics of functions $Y$ and $L$, we obtain the following inequality 
\begin{align}
L(T)-Y(T)&=\int^T_{t_0} r(L(s)-Y(s))ds+\int^T_{t_0}(\pi^{2,\eps}_s-C_2)ds\nonumber\\
&\geq \int^T_{t_0}(\pi^{2,\eps}_s-C_2)ds\geq 0\label{last}.
\end{align}
From (\ref{KK}), we obtain $L(T)\rightarrow S$, as $\eps \rightarrow 0$. So, the minimum guarantee constraint, $X(T;t_0, x_0, \pi^{1,\eps}(\cdot), \pi^{2,\eps}(\cdot))\geq S \; a. s.$, and consequently $Y(T)\geq S$, implies that the left hand side tends to zero which concludes the convergence (\ref{consumptionconver}).  

Now, (\ref{first}) and (\ref{two}) imply the claim, 
$$\tilde{V}(t_0, S(t_0)+\eps) \rightarrow \tilde{V}(t_0, S(t_0)), \qquad as \; \eps \rightarrow 0. $$\\
 
\emph{Step 5.} The two variable function $(t,x) \rightarrow V(t,x)$ is continuous at the boundary $[0, T]\times \{S\}$. 

\end{document}